\documentclass[a4paper,fleqn]{cas-dc}

\usepackage[authoryear]{natbib}
\usepackage{caption}
\usepackage{siunitx}
\usepackage{graphicx}
\usepackage{ulem}
\newcommand{\R}[1]{{\textcolor{black}{#1}}}

\def\tsc#1{\csdef{#1}{\textsc{\lowercase{#1}}\xspace}}
\tsc{WGM}
\tsc{QE}

\begin{document}
\let\WriteBookmarks\relax
\def\floatpagepagefraction{1}
\def\textpagefraction{.001}

\shorttitle{}    

\shortauthors{}  

\title [mode = title]{Flow birefringence measurement in a radial Hele-Shaw cell considering three-dimensional effects}  



%

\author[1,2]{Misa Kawaguchi}[orcid=0009-0001-8637-5422]



\ead{kawaguchi@shinshu-u.ac.jp}


\credit{Conceptualization, Methodology, Funding acquisition, Writing - Original Draft}

\affiliation[1]{organization={Institute of Engineering, Academic Assembly, Shinshu University},
            postcode={380-8553}, 
            state={Nagano},
            country={Japan}}

\author[2]{William Kai Alexander Worby}



\credit{Methodology, Validation, Writing – review \& editing}

\affiliation[2]{organization={Department of Mechanical Systems Engineering, Tokyo University of Agriculture and Technology},
            postcode={184-8588}, 
            state={Tokyo},
            country={Japan}}



\author[3]{Yuto Yokoyama}

\credit{Conceptualization, Methodology, Writing – review \& editing}
\affiliation[3]{organization={Micro/Bio/Nanofluidics Unit, Okinawa Institute of Science and Technology},
            postcode={904-0495}, 
            state={Okinawa},
            country={Japan}}
            
\author[4]{Ryuta X. Suzuki}

\credit{Methodology, Resources, Funding acquisition, Writing – review \& editing}
\affiliation[4]{organization={West Tokyo Joint Center for Sustainability Research and Implementation, Tokyo University of Agriculture and Technology},
            postcode={184-8588}, 
            state={Tokyo},
            country={Japan}}
            
\author[5]{Yuichiro Nagatsu}

\credit{Conceptualization, Resources, Methodology, Writing – review \& editing}
\affiliation[5]{organization={Department of Chemical Engineering, Tokyo University of Agriculture and Technology},
            postcode={184-8588}, 
            state={Tokyo},
            country={Japan}}
            
\author[2,6]{Yoshiyuki Tagawa}

\credit{Conceptualization, Methodology, Resources, Funding acquisition, Writing – review \& editing}
\affiliation[6]{organization={Institute of Global Innovation Research, Tokyo University of Agriculture and Technology},
            postcode={184-8588}, 
            state={Tokyo},
            country={Japan}}

\cormark[1]


\ead{tagawayo@cc.tuat.ac.jp}


\begin{abstract}
Flow birefringence measurement is an emerging technique for visualizing stress fields in fluid flows. This study investigates flow birefringence in the steady radial Hele-Shaw flow. In the radial Hele-Shaw flow, stress is dominant along the gap direction, challenging the applicability of the conventional stress-optic law (SOL) with measurement from the gap direction. To overcome this problem, we used two types of flow birefringence measurement using radial Hele-Shaw cell and rheometer. We conduct flow birefringence measurements at various flow rates and compare the results with theoretical predictions. The observed phase retardation cannot be quantitatively explained using the conventional SOL, but is successfully described using the second-order SOL, which accounts for stress along the optical direction. The stress-optic coefficient in the second-order SOL was obtained by rheo-optical measurements. This study demonstrates that the combination of the second-order SOL and rheo-optical measurements is essential for an accurate interpretation of flow birefringence in Hele-Shaw flow, providing a noninvasive approach for stress field analysis in high-aspect-ratio geometries.
\end{abstract}



\maketitle

\section{Introduction}\label{intro}
One example of confined thin flow is Hele-Shaw flow, which occurs in the thin space between two parallel plates. Hele-Shaw flow can be used to visualize the streamlines of two-dimensional Euler flows \citep{hele1898flow,batchelor2000introduction}, to study the Saffman--Taylor instability \citep{homsy1987viscous,saffman1958penetration}, to examine biological phenomena such as cell adhesion \citep{rezania1997probabilistic,goldstein1998comparison,delon2020hele}, and to evaluate cleaning processes \citep{deshmukh2022cleaning}.
Thus, the measurement of the stress field in a Hele-Shaw cell is important in understanding mechanobiology, the fluid mechanics of instability phenomena, and cleaning processes. 

The conventional approach for obtaining the stress field is to obtain velocity measurements, such as through particle image velocimetry \citep{kloosterman2011flow,ehyaei2014quantitative}. However, predicting the shear stresses in thin flow regions remains challenging. Other promising technique is flow birefringence measurement.  Suspensions of crystal-like particles or polymers generate flow birefringence (double refraction) when a velocity gradient is applied \citep{maxwell1873double}.
Flow birefringence is a useful tool for visualizing flow patterns and understanding the rheology of complex fluids. In the absence of flow, the suspension is optically isotropic because nonspherical particles or polymers are randomly oriented due to Brownian motion. When flow is applied, local velocity gradients induce particle orientation and polymer stretch, and this leads to an optically anisotropic state in the fluid. These properties are exploited for flow visualization in biomechanics \citep{rankin1989streaming,sun1999visualisation,sun2016measurements} and biolocomotion \citep{hu2009flow}. Using the optical anisotropy induced by the alignment of suspended particles, flow birefringence has been applied to rheo-optic measurements \citep{ober2011spatially,sato2024two, worby2024examination}, which enable the study of macromolecules \citep{cerf1952flow}. 

Flow birefringence can also be used in the experimental determination of shear rate and stresses based on the stress-optic law (SOL) \citep{aben2012photoelasticity}, which expresses the relationship between birefringence and stresses. 
Studies of flow birefringence using the SOL have mostly considered two-dimensional flow \citep{noto2020applicability} or uniform shear flow \citep{decruppe2003flow,lane2022birefringent,worby2024examination}. However, there are several unsolved problems regarding flow birefringence measurements in a Hele-Shaw cell, such as the shear rate and stress field evaluation. First, the dominant components of stress in Hele-Shaw flow are not included in the conventional SOL when the optical axis is axial direction in cylindrical coordinate. The conventional SOL is not sufficient to understand the three-dimensional phenomena that occur in such as Hele-Shaw flow and rectangular channels. \citet{nakamine2024flow} suggested that the second-order SOL (explained in \S\ref{subsec2.1}) has the potential to explain the effects of stress along the optical axis. 

The purpose of this study is to investigate flow birefringence in steady radial Hele-Shaw flow, considering the effects of stress along the optical axis.
We conduct flow birefringence measurements using cellulose nanocrystal (CNC) suspensions over a range of flow rates, and compare the results with theoretical predictions based on the second-order SOL. Radial Hele-Shaw flow generates a spatially dependent shear stress, making it suitable for validation of the integrated photoelasticity. Here, we derive the second-order SOL for Hele–Shaw flow, where the dominant stress components differ from those in previous studies, such as rectangular channel flow and simple shear flow in a rheometer. The radial Hele-Shaw flow provides a canonical geometry for a simple flow exhibiting variation of shear stress along the optical axis.
\R{Building on our previous work \citep{nakamine2024flow}, where the necessity of the second-order SOL was demonstrated and the coefficient $C_2$ was determined through fitting, the present study advances the approach by (i) independently calibrating $C_2$ via rheo-optical measurements, (ii) applying the calibrated $C_2$ to radial Hele-Shaw flow, and (iii) validating the framework in a path-integrated radial Hele-Shaw geometry, involving stress variations along both the optical-axis and radial directions, rather than in rectangular or simple shear configurations.}

The remainder of this paper is organized as follows. In \S 2, the principle of the photoelastic technique and the theoretical solution of phase retardation, which is obtained by integrating the birefringence along the optical path, in radial Hele-Shaw flow are introduced. In \S 3, the experimental methods are presented. In \S 4, the results of phase retardation in Hele-Shaw flow are presented and the applicability of the SOL considering three-dimensional stress fields is discussed. Finally, the conclusions from this study are summarized in \S 5.

\section{Calculation of theoretical phase retardation}\label{calc}

\S\ref{subsec2.2} describes the velocity distribution and stress tensor in Hele-Shaw flow. In \S \ref{subsec2.1} explains principle of photoelastic measurement. 
In \S \ref{subsec2.3}, we identify the relationship between the stress components and the phase retardation. 


\subsection{Radial Hele-Shaw flow}\label{subsec2.2}
\begin{figure}
\centering
\includegraphics[width=1\linewidth]{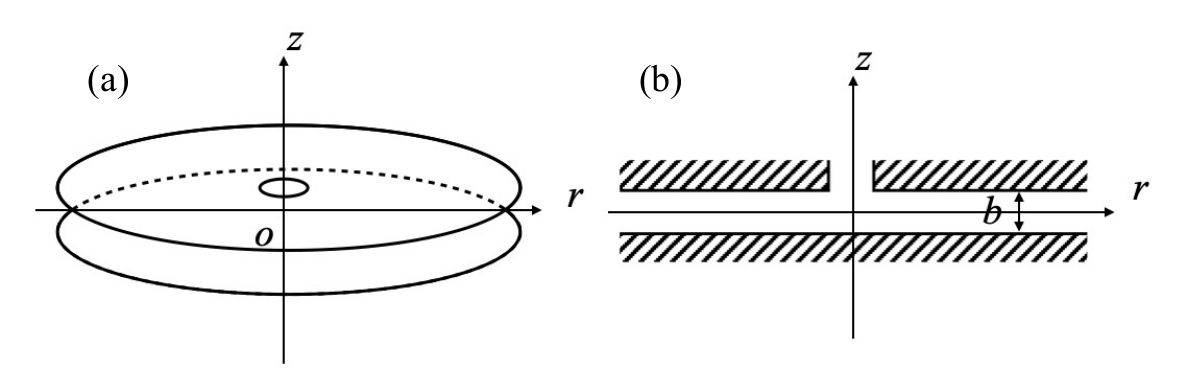}
\caption{Configuration and coordinates in radial Hele-Shaw flow}
\label{fig:config}
\end{figure}

 The configuration and coordinates are shown in Fig. \ref{fig:config}. Two parallel plates are separated by a gap $b$. In this study, the only injection point is on the upper side. 

The continuity and momentum equations reduce to the following forms with the assumption of radial flow ($u_\theta$ and $u_z$ are 0) and incompressible and steady flow in radial Hele-Shaw cell.
\begin{align}
\textcolor{black}{
\frac{\partial{u_r}}{\partial{r}}+\frac{u_r}{r}
}&\textcolor{black}{=0,}
\label{eq:cont}\\
\textcolor{black}{\rho u_r\frac{\partial{u_r}}{\partial{r}}}
&\textcolor{black}{{=-\frac{dp}{dr}
+\mu\left(\frac{\partial^2{u_r}}{\partial{r^2}}
+\frac{\partial{u_r}}{r\partial r}
-\frac{u_r}{r^2}
+\frac{\partial^2{u_r}}{\partial z^2}\right),}
}
\label{eq:NS}
\end{align}
where $\rho$ is the fluid density and $\mu$ is the viscosity. 

The continuity equation (Eq.\eqref{eq:cont}) substitute to the momentum equation (Eq. \eqref{eq:NS}), then the equation of motion reduces as follows:
\begin{align}
\textcolor{black}{
\rho\frac{{u_r}^2}{r} = -\frac{dp}{dr}+\mu\frac{\partial^2{u_r}}{\partial z^2}.
}
\end{align}

The velocity profile can be expressed as

\begin{align}
u_r(r,z) &= 
\frac{3Q}{4\pi rb} 
\left\{
1-\left(\frac{2z}{b}\right)^{2}
\right\} ,
\label{eq:HS_vel}
\end{align}
where $Q$ is the flow rate \R{and $b$ is the gap of the Hele-Shaw cell}. 

The deformation rate tensor $\mathbf{D}$ and the viscous stress tensor $\mathbf{T_{HS}}$ in radial Hele-Shaw flow are derived under the assumptions of axial symmetry around the $z$-axis, $u_{\theta}=0$, and $u_z=0$:

\begin{equation}
\begin{aligned}
\mathbf{D} &= \frac{1}{2} \left( \nabla \mathbf{u} + \nabla \mathbf{u}^\top \right) 
=
\begin{pmatrix}
\frac{\partial u_r}{\partial r} & 0 & \frac{1}{2} \left( \frac{\partial u_r}{\partial z} \right) \\[10pt]
0 & \frac{u_r}{r} & 0 \\[10pt]
\frac{1}{2} \left( \frac{\partial u_r}{\partial z} \right) & 0 & 0
\end{pmatrix},
\end{aligned}
\label{eq:defotmation_tensor}
\end{equation}
\begin{equation}
\begin{aligned}
\mathbf{T_{HS}} &= 2\mu \mathbf{D} \\
&=
\begin{pmatrix}
  \textcolor{black}{2\mu\frac{\partial u_r}{\partial r}} & 0 & \mu\frac{\partial u_r}{\partial z} \\
0 & \textcolor{black}{2\mu\frac{u_r}{r}} & 0 \\
\mu\frac{\partial u_r}{\partial z} & 0 & \textcolor{black}{0} \\
\end{pmatrix}.
\end{aligned}
\label{eq:stress_tensor}
\end{equation}

From Eqs.\eqref{eq:HS_vel} and \eqref{eq:stress_tensor}, the viscous stress components are
\begin{equation}
\begin{aligned}
\textcolor{black}{\sigma_{rr}} &\textcolor{black}{= -\frac{3\mu Q}{2\pi r^2b}
\left\{1-\left(\frac{2z}{b} \right)^{2}\right\},}\\
\textcolor{black}{\sigma_{\theta \theta}} &\textcolor{black}{= \frac{3\mu Q}{2\pi r^2b}
\left\{1-\left(\frac{2z}{b} \right)^{2}\right\}
=-\sigma_{rr},}\\
\textcolor{black}{\sigma_{rz}} &\textcolor{black}{= -\frac{6\mu Qz}{\pi rb^3},}\\
\end{aligned}
\label{eq:stress_components}
\end{equation}
under the assumption of axis symmetry.

\subsection{Principle of photoelastic measurement}\label{subsec2.1}
The quantitative evaluation of flow birefringence requires the second-order SOL to be considered \citep{nakamine2024flow}. 
The second-order SOL relates all components of the stress tensor with the phase retardation, while the first-order SOL neglects the stresses along the optical axis.
The retardation $\Delta$ and orientation $\phi$ is related to the components of dielectric tensor as follows (\cite{aben1997photoelastic}):

\begin{align}
V_1 
&=\Delta \cos2\phi\nonumber\\
&=\frac{1}{2n_0}\int\{(\epsilon_{\theta \theta}-\epsilon_{rr})\cos2\theta\nonumber\\&~~~~~~~~~~~~~~~~~~~+2\epsilon_{r\theta}\sin2\theta\}dz,
\label{eq:v1_eps}
\end{align}

\begin{align}
V_2 &=\Delta \sin2\phi\nonumber\\
&=\frac{1}{n_0}\int\{(\epsilon_{rr}-\epsilon_{\theta\theta})\sin\theta\cos\theta\nonumber\\&~~~~~~~~~~~~~~~~+\epsilon_{r \theta }\cos2\theta\} dz.
\label{eq:v2_eps}
\end{align}
where $\epsilon_{ij}$ is the components of dielectric tensor $\bf{\epsilon}$ in cylindrical coordinate with an assumption of axis symmetry. Note that the optical axis is set to the $z$-direction. $\Delta$ can be expressed as

\begin{equation}
\Delta = \sqrt{V_1^2+V_2^2}\label{eq:delta}.
\end{equation}
This is valid under the condition where retardation is less than approximately quarter of the wavelength and the angle of rotation of the principal directions does not exceed $\pi$/6 (\cite{aben1997photoelastic}).

\R{The optical effect is a function of the strain rates $\dot{e}_{ij}$ and it can be written using Cayley–Hamilton theorem as \citep{aben1997photoelastic, doyle1982nonlinearity}}

\begin{align}
\frac{1}{2n_0}\epsilon_{ij} &= f(\dot{e}_{ij})\nonumber\\
&=\alpha_0\delta_{ij}+\alpha_1\dot{e}_{ij}+\alpha_2\dot{e}_{ik}\dot{e}_{kj},
\label{eq:Cayley–Hamilton}
\end{align}

\R{where $\alpha_i$ depends on the fluid and of the invariants from $\dot{e}_{ij}$.}

\R{Eq. \eqref{eq:Cayley–Hamilton} reveals \citep{aben1997photoelastic}}

\begin{align}
\frac{1}{2n_0}\{(\epsilon_{\theta \theta}-\epsilon_{rr})\cos2\theta+2\epsilon_{r\theta}\sin2\theta \}\nonumber\\
=\alpha_1\bigl[({\dot{e}_{\theta \theta}} - {\dot{e}_{rr}})\cos2\theta\nonumber+2\dot{e}_{r\theta}\sin2\theta\bigr]\nonumber\\
+\alpha_2\bigl[({\dot{e}_{rr}} + {\dot{e}_{\theta \theta}})\{({\dot{e}_{\theta \theta}} - {\dot{e}_{rr}})\cos2\theta\nonumber +2\dot{e}_{r\theta}\sin2\theta\}  \nonumber \\+ (\dot{e}_{rz}\sin\theta + \dot{e}_{\theta z}\cos\theta)^2  ({\dot{e}_{rz}\cos\theta-\dot{e}_{\theta z}\sin\theta})^2\bigr],\\
\frac{1}{n_0}\{(\epsilon_{rr}-\epsilon_{\theta\theta})\sin\theta\cos\theta+\epsilon_{r \theta }\cos2\theta\ \} \nonumber\\
=2{\alpha_1}\bigl[\dot{e}_{rz}\sin\theta + \dot{e}_{\theta z}\cos\theta\bigr] \nonumber \\
+{\alpha_2}\bigl[({\dot{e}_{rr}} + {\dot{e}_{\theta \theta}})\{({\dot{e}_{\theta \theta}} - {\dot{e}_{rr}})\cos2\theta\nonumber
+2\dot{e}_{r\theta}\sin2\theta\}  \nonumber \\+2(\dot{e}_{rz}\sin\theta + \dot{e}_{\theta z}\cos\theta)
(\dot{e}_{rz}\cos\theta-\dot{e}_{\theta z}\sin\theta)\bigr].
\end{align}

\R{Then, the equations \eqref{eq:v1_eps} and \eqref{eq:v2_eps} can be written as}

\begin{align}
V_1 &=\int \Bigl(\alpha_1\bigl[({\dot{e}_{\theta \theta}} - {\dot{e}_{rr}})\cos2\theta\nonumber+2\dot{e}_{r\theta}\sin2\theta\bigr] \nonumber \\
&~~~~~~~~~~~~+ \alpha_2\bigl[({\dot{e}_{rr}} + {\dot{e}_{\theta \theta}})\{({\dot{e}_{\theta \theta}} - {\dot{e}_{rr}})\cos2\theta\nonumber \\&~~~~~~~~~~~~~~~~~~~~~~~~~~~~~~~~~~~~~~~~~+2\dot{e}_{r\theta}\sin2\theta\}  \nonumber \\&~~~~~~~~~~~~~~~~~~~~~~~+ (\dot{e}_{rz}\sin\theta + \dot{e}_{\theta z}\cos\theta)^2 \nonumber \\
&~~~~~~~~~~~~~~~~~~~~~~~- ({\dot{e}_{rz}\cos\theta-\dot{e}_{\theta z}\sin\theta})^2\bigr]\Bigr) \, dz,
\label{eq:v1_srate}
\end{align}

\begin{align}
V_2 &= \int \Bigl( 2{\alpha_1}\bigl[\dot{e}_{rz}\sin\theta + \dot{e}_{\theta z}\cos\theta\bigr] \nonumber \\
&~~~~~~~~~~~+{\alpha_2}\bigl[({\dot{e}_{rr}} + {\dot{e}_{\theta \theta}})\{({\dot{e}_{\theta \theta}} - {\dot{e}_{rr}})\cos2\theta\nonumber \\
&~~~~~~~~~~~~~~~~~~~~~~~~~~~~~~~~~~~~~~~~+2\dot{e}_{r\theta}\sin2\theta\}  \nonumber \\&~~~~~~~~~~~~~~~~~~~~~+2(\dot{e}_{rz}\sin\theta + \dot{e}_{\theta z}\cos\theta)\nonumber \\&~~~~~~~~~~~~~~~~~~~~~~~~~~~~(\dot{e}_{rz}\cos\theta-\dot{e}_{\theta z}\sin\theta)\bigr] \Bigr)dz.
\label{eq:v2_srate}
\end{align}
where $\alpha_1$ and $\alpha_2$ are the function of physical properties of test fluid and of the invariants from strain rates $\dot{e}_{ij}$ in cylindrical coordinate. Note that this assumption for $\alpha_1$ and $\alpha_2$ has not been validated either theoretically or experimentally.
\cite{nakamine2024flow} extended Eqs. \eqref{eq:v1_srate} and \eqref{eq:v2_srate} for Newtonian fluid using viscous stress, which is proportional to the strain rate, then

\begin{align}
V_1 &=\int \Bigl(C_1\bigl[({\sigma_{\theta \theta}} - {\sigma_{rr}})\cos2\theta\nonumber +2\sigma_{r\theta}\sin2\theta\bigr] \nonumber \\
&~~~~~~~~~~~~+ C_2\bigl[({\sigma_{rr}} + {\sigma_{\theta \theta}})\{({\sigma_{\theta \theta}} - {\sigma_{rr}})\cos2\theta\nonumber \\&~~~~~~~~~~~~~~~~~~~~~~~~~~~~~~~~~~~~~~~~~+2\sigma_{r\theta}\sin2\theta\}  \nonumber \\&~~~~~~~~~~~~~~~~~~~~~~~+ (\sigma_{rz}\sin\theta + \sigma_{\theta z}\cos\theta)^2 \nonumber \\
&~~~~~~~~~~~~~~~~~~~~~~~- ({\sigma_{rz}\cos\theta-\sigma_{\theta z}\sin\theta})^2\bigr]\Bigr) \, dz,
\label{eq:v1}
\end{align}

\begin{align}
V_2 &= \int \Bigl( 2{C_1}\bigl[\sigma_{rz}\sin\theta + \sigma_{\theta z}\cos\theta\bigr] \nonumber \\
&~~~~~~~~~~~+{C_2}\bigl[({\sigma_{rr}} + {\sigma_{\theta \theta}})\{({\sigma_{\theta \theta}} - {\sigma_{rr}})\cos2\theta\nonumber \\
&~~~~~~~~~~~~~~~~~~~~~~~~~~~~~~~~~~~~~~~~+2\sigma_{r\theta}\sin2\theta\}  \nonumber \\&~~~~~~~~~~~~~~~~~~~~~+2(\sigma_{rz}\sin\theta + \sigma_{\theta z}\cos\theta)\nonumber \\&~~~~~~~~~~~~~~~~~~~~~~~~~~~~(\sigma_{rz}\cos\theta-\sigma_{\theta z}\sin\theta)\bigr] \Bigr)dz.
\label{eq:v2}
\end{align}
where $C_1$ [Pa$^{-1}$] and $C_2$ [Pa$^{-2}$] are the stress-optic coefficients, and $\sigma_{ij}$ denotes the stress components in cylindri-
cal coordinate. 
When $C_2 = 0$, Eqs. \eqref{eq:delta}, \eqref{eq:v1} and \eqref{eq:v2} correspond to the conventional SOL \citep{prabhakaran1975stress} used in solid mechanics.

The phase retardation $\Delta$ is obtained from a four-step phase-shifting method \citep{ramesh2021developments,otani1994two,onuma2014development}. 
The retardation is calculated as follows using four intensities captured by super-pixels of the image sensor:
\begin{equation}
\Delta = \frac{\lambda}{2\pi}\sin^{-1}\frac{\sqrt{(I'_{90}-I'_0)^2 + (I'_{45} - I'_{135})^2}}{I/2},
\label{eq:retardation}
\end{equation}
where $\lambda$ is the wavelength of the light and $I'$ is the intensity detected by the camera sensor through a polarizer; the subscript denotes the angle of the polarizer.
$I$ is the intensity of the incident light, given by
\begin{equation}
I = I'_0 + I'_{45} + I'_{90} + I'_{135}.
\end{equation}
Note that the Eq. \eqref{eq:retardation} is valid where the retardation is less than a quarter of the wavelength of the light ($\pi$/2 rad) (\cite{onuma2014development}).

\subsection{Theoretical solution of phase retardation}\label{subsec2.3}
The theoretical solution of phase retardation for $\theta=0$ is calculated based on Eqs. \eqref{eq:delta}, \eqref{eq:v1} and \eqref{eq:v2} using the stress components in Eq. \eqref{eq:stress_components}:

\begin{equation}
\begin{aligned}
\textcolor{black}{\Delta} &\textcolor{black}{= \sqrt{V_1^2+V_2^2} = \left|V_1\right|} \\
&\textcolor{black}{= \int_{-b/2}^{b/2} \bigl(2C_1{\sigma_{\theta \theta}} - C_2{\sigma_{rz}}^2\bigr)dz.}\\
&\textcolor{black}{= \int_{-b/2}^{b/2} \Bigg[\frac{6C_1\mu Q}{2\pi r^2b}
\left(1-\left|\frac{2z}{b} \right|^{2}\right)}\\
&\quad
\textcolor{black}{-C_2\left(\frac{3\mu Q}{2\pi rb}
\left|\frac{2z}{b} \right| \right)^2 \Bigg]dz.}
\end{aligned}
\label{eq:delta_analytical}
\end{equation}

For the stress-optic coefficients $C_1$ and $C_2$, \cite{nakamine2024flow} explained their experimental results by assuming the values corresponding to each condition, where $C_1$ is a function of CNC concentration and $C_2$ is a value that depends on CNC concentration and channel aspect ratio. For $C_2$, \cite{worby2024examination} reported that $C_2$ is a power function of the second invariant of the deformation-rate tensor. Based on this observation, we performed the calibration of $C_2$ for Eq. \eqref{eq:delta_analytical} by rheo-optical measurement (see \S \ref{RO_measure}). Note that the reason $C_2$ depends on the shear rate is an unsolved problem.

Here, we assume that $C_1$ equals $1 \times 10^{-5}$ [Pa$^{-1}$]. This is a rather bold assumption, but in a Hele-Shaw flow, the way to determine $C_1$ contributes little to the magnitude of phase retardation (Appendix \ref{appB}). The contribution of the second term on the left-hand side of Eq. \eqref{eq:v1} (i.e., the way of determination of $C_2$) is dominant.

\section{Experimental methods}\label{exp}
\subsection{Photoelastic measurement of radial Hele-Shaw flow}\label{emethod_HS}

\begin{figure}
\centering
\includegraphics[width=0.95\linewidth]{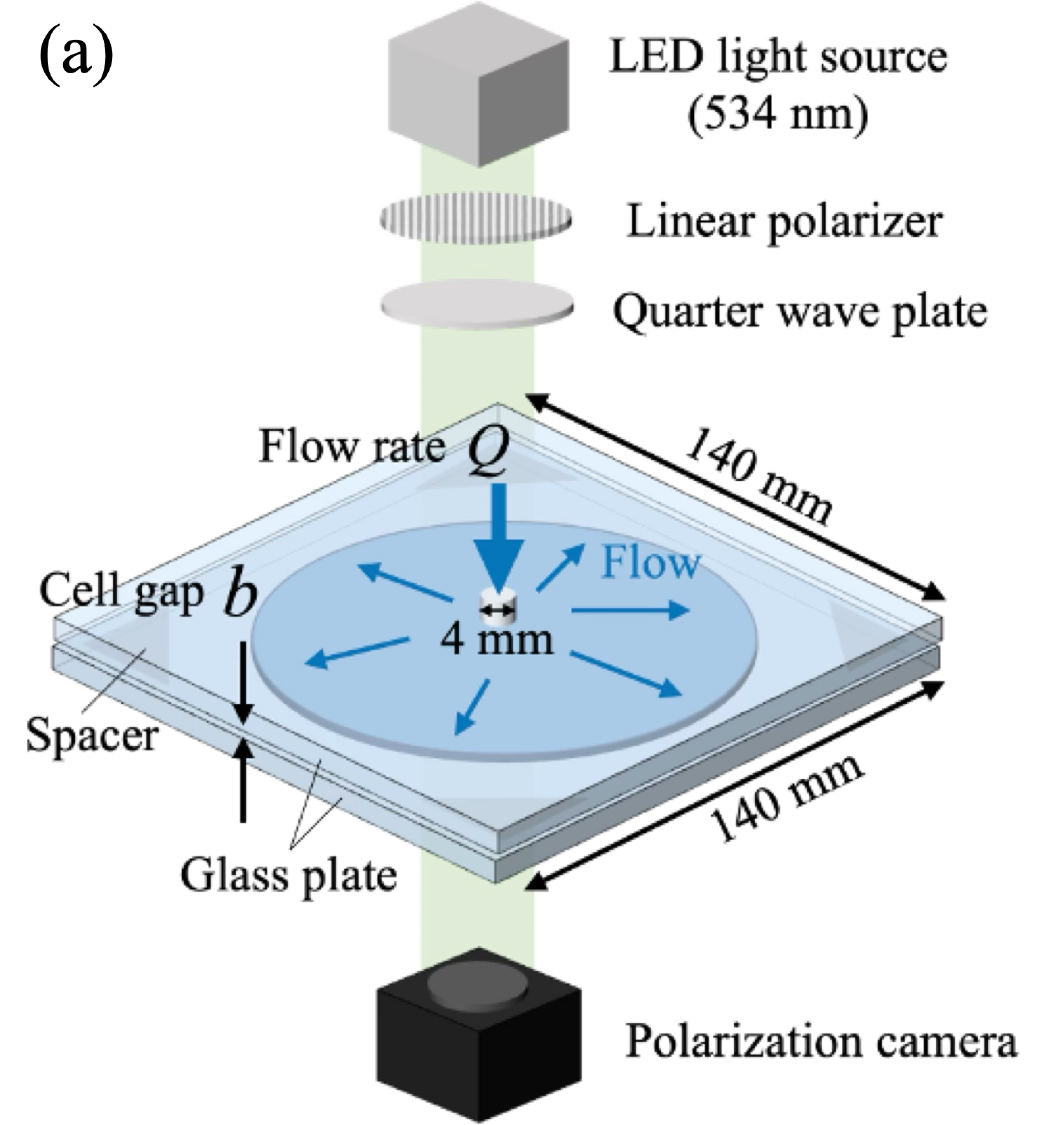} 
\includegraphics[width=0.9\linewidth]{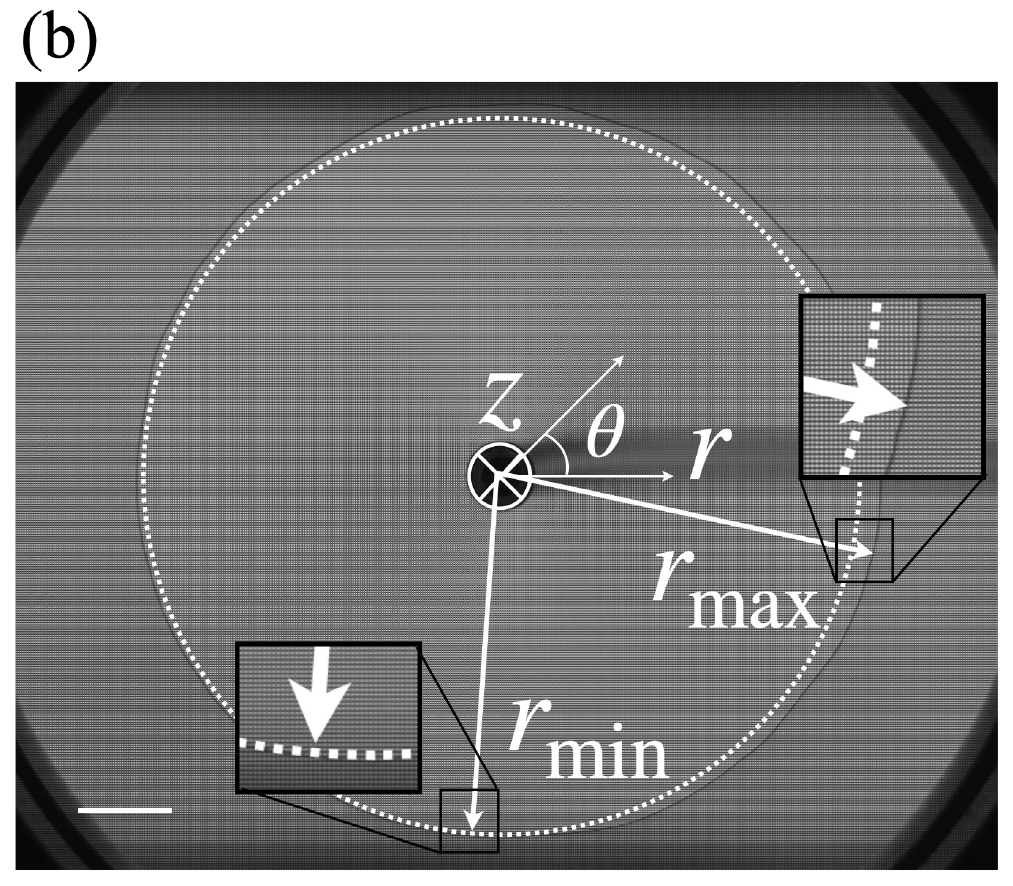} 
\caption{(a)Experimental setup using a Hele-Shaw cell. (b) Example of raw image. The white dotted circle has $r$ = 35 mm, the radius at the furthest interface from the center of inlet is defined as $r_\mathrm{max}$ and the radius at the nearest interface from the center of inlet is defined as $r_\mathrm{min}$. The white bar indicates a scale of 10 mm.}
\label{fig:setup}
\end{figure}

The experimental setup for the photoelastic measurements of radial Hele-Shaw flow is shown in Fig. \ref{fig:setup}. The radial Hele-Shaw cell consists of two soda-lime glass plates (140 mm $\times$ 140 mm, Furukawa Science and Engineering). The upper glass plate has a small hole (4 mm diameter) in its center for liquid injection. These two glasses are separated by four triangular metal plates (40 mm $\times$ 40 mm) placed at the four corners. The thickness of the metal spacers, as measured by a micrometer (Mitutoyo, Digimatic micrometer, MPC-25MX) was $0.283\pm0.012$ mm. These measurements were taken at three different locations for each spacer. The glass plates and spacers of the Hele-Shaw cell were fixed with a flange. 
The CNC (cellulose nanocrystal) suspension was injected from the upper side of the Hele-Shaw cell using a syringe pump (YSP-301(450258), YMC) with a constant flow rate $Q$, where $Q$ was varied from 25--50 ml/min at intervals of 5 ml/min. This condition was adopted as the condition under which the interface spreads while maintaining a relatively circular shape. Three measurements were conducted for each condition to ensure reliability and reproducibility. 
Circularly polarized light from a light source with wavelength $\lambda$ = 543 nm (SOLIS-565C, Thorlabs) passed through the linear polarizer and a quarter-wave plate was emitted through the flow field.
Elliptically polarized light with retardation and azimuth was then captured by a polarization camera (CRYSTA PI-5WP, Photron) placed under the Hele-Shaw cell at a frame rate of 250 fps and spatial resolution of 226 \textmu m/pixel. \R{The measurement range is limited in our setup by the spatial resolution and super-pixel ($\sim$ 450 \textmu m) and the field of view (95.8 mm $\times$ 76.8 mm), corresponding to mesoscopic length scales ($\sim 5\times10^2$ \textmu m to ~$10^1$ mm).} The photoelastic measurements were performed at room temperature of 22${}^\circ$C.
The phase retardation was calculated from the intensity using Eq. \eqref{eq:retardation} in the CRYSTA Stress Viewer software (Photron).

To validate the photoelastic measurements, the phase retardation of a polarizer with known retardations of 50.8 nm and 126.2 nm was measured. 
The mean and standard deviation of the measured phase retardation were 48.3$\pm$1.35 nm for polarizer with a known value of 50.8 nm, and 127.6$\pm$1.75 nm for polarizer with a known value of 126.2 nm.

\subsection{Suspension preparation}\label{emethod_cnc}
The CNC (cellulose nanocrystal high sulfonic group content, freeze‐dried (CNC-HSFD), Cellulose Lab) was dispersed in ultrapure water (Milli-Q, Merck) as birefringent material. There are many birefringent materials, such as tobacco mosaic virus \citep{hu2009flow}, milling yellow \citep{pindera1978characteristic}, CNC \citep{lane2024two}, and KaleidoFlow \citep{noto2020applicability}. Our group examined several birefringent materials, and selected CNC-HSFD as a high-sensitivity example from among the materials currently available. 

The concentration of the CNC was set to 3 wt\%. In this condition, concentration regime is classified to semi-dilute regime by Doi-Edwords theory (\cite{doi1988theory}). The average length $l$ and diameter $d$ of CNC are 175 nm and 12.5 nm, respectively (catalog by Cellulose lab), and the volume concentration $\phi_{vol}$ is 2.0\% based on the density of CNC (1,500 kg/m$^3$, Safety Data Sheet from Cellulose lab) and that of water (998 kg/m$^3$). Then, the number density $n$ = 4$\phi_{vol}$ /($\pi ld^2$) becomes $9.4\times10^{20}$. \cite{doi1988theory} define a concentration regime for rigid rod-like polymers: the condition $n \lesssim n_1 \simeq 1/l^3$ means dilute, and the condition $n_1 \lesssim n \lesssim n_2 \simeq 1/dl^2$ is semi-dilute regime, while the condition $n_2 \lesssim n$ indicates concentrated regime. According to their definition, $n_1 = 1.0\times10^{20}$ and $n_2= 4.3\times10^{21}$ in our condition. Therefore, our condition is classified semi-dilute regime because of high aspect ratio. If the CNC concentration is less than 3 wt\% under our conditions, the flow birefringence becomes too weak to detect and is hidden within the noise.

The CNC suspension was sonicated using a homogenizer (UX-300, Mitsui Electric) to ensure that the particles were homogeneously dispersed. Ultrasound treatment was applied for 400 s (on time: 5 s, off time: 10 s, set time: 200 s $\times$ two sets) at 40 \% power in an ice bath to prevent overheating.

\begin{figure}
\centering
\includegraphics[width=1\linewidth]{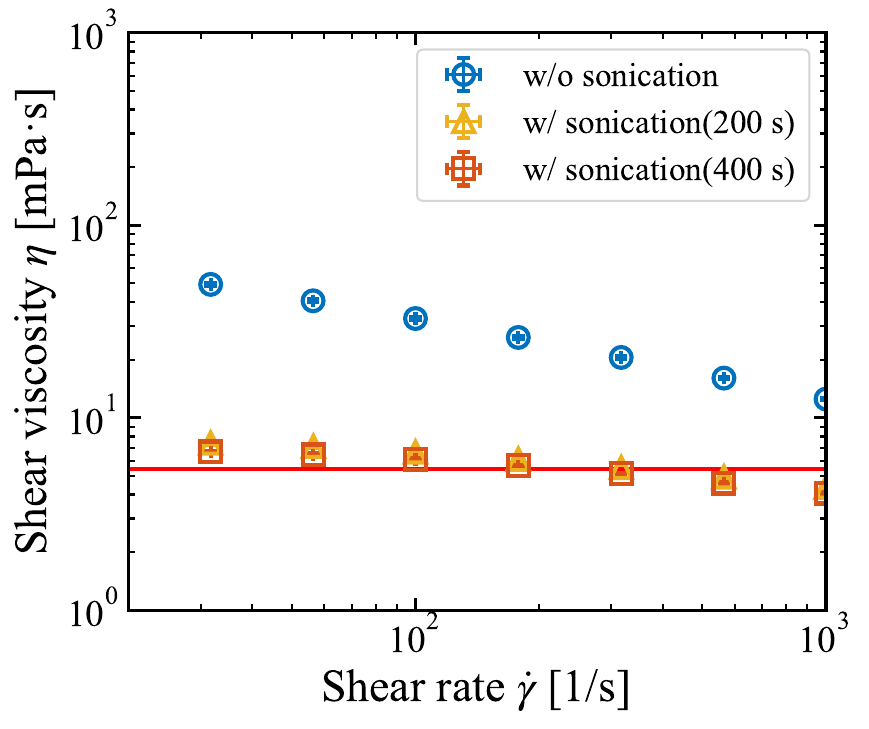}
\captionsetup{justification=justified}
\caption{Shear viscosity measured by rheometer at ${20}^\circ$C; \R{Error bar indicates $\pm$1SD from three independent measurements (n = 3).} Blue circles: No sonication; Yellow triangles: Sonication for 200 s; Red squares: Sonication for 400 s. Red solid line is the weighted average value for 10$<\dot{\gamma}<$700.}
\label{fig:shear_vis}
\end{figure}

The shear viscosity was measured by a rheometer (MCR302, Anton Paar) equipped with a cone-plate (CP50-0.5, Anton Paar) at 20${}^\circ$C (Fig. \ref{fig:shear_vis}), as controlled by a temperature system in the rheometer. Before sonication, the CNC suspension exhibited pronounced shear-thinning due to the CNC particles. After sonication, the shear-thinning of the CNC suspension was suppressed and the shear viscosity had decreased, as observed in a previous study \citep{shafiei2012rheology}.

In this study, the viscosity $\mu$ for calculation of theoretical solution (Eq. \eqref{eq:delta_analytical}) is assumed to be constant under the assumption of a Newtonian fluid for the working fluid. To avoid bias due to nonuniform sampling in shear rate $\dot{\gamma}$, the average value of shear viscosity $\eta(\dot{\gamma})$ was evaluated as an $\dot{\gamma}$-weighted mean using data obtained after 400 s of sonication, over the shear-rate range 10$<\dot{\gamma}<$700, which corresponds to 25<$Q$<50 ml/min and 15<$r$<35 mm in the radial Hele-Shaw flow experiment. The approximation of Newtonian fluid includes from $-11\%$ to $36\%$ error. \R{The effects of shear-thinning on the theoretical solution is presented in Appendix \ref{app_non-newtonian}.}

\subsection{Calibration measurement of $C_2$}\label{RO_measure}
\begin{figure}
\centering
\includegraphics[width=1\linewidth]{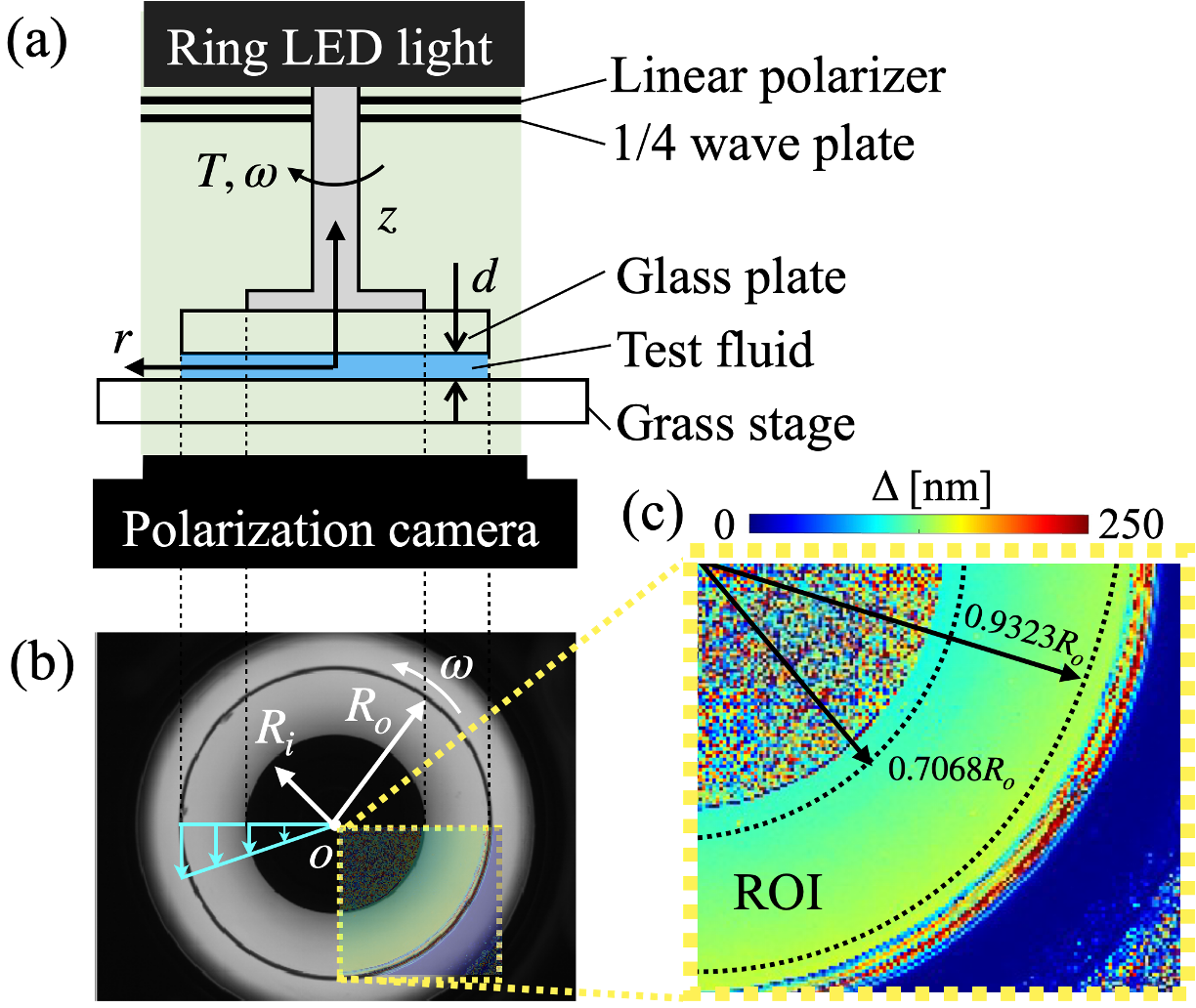}
\captionsetup{justification=justified}
\caption{(a) Experimental setup for rheo-optical measurements, consisting of high-speed polarization camera and a rheometer with a glass parallel plate and stage. $T$ is torque, $\omega$ is angular velocity. Plate diameter is $2R_o$ = 43.016 mm. (b) Example of raw image and (c) diagram of a quarter area of phase retardation distribution for $\omega$ = 11.2 rad/s. Region of interest (ROI) is shown. The phase retardation in the ROI was used to calculate $C_2$ (Eq. \eqref{eq:C2_calc}).}
\label{fig:rheometer}
\end{figure}

To evaluate the stress-optic coefficient $C_2$, photoelastic measurements were performed using a rheometer. The experimental setup is shown in Fig. \ref{fig:rheometer}. A parallel-plate type rheometer (MCR302, Anton Paar) with a transparent plate (PP43/GL-HT, Anton Paar) and a stage (PTD200/GL, Anton Paar) were used, with the gap $d$ set to 0.3 mm. A ring-shaped light source with wavelength $\lambda$ = 543 nm (LDR70SE2, CCS) was mounted on the rheometer. The light passed through the linear polarizer and a quarter-wave plate was emitted through the flow field. 
The phase retardation was measured by a polarization camera (CRYSTA PI-5WP, Photron) placed under the rheometer at a frame rate of 250 fps in the steady state. The measured data from 100 frames were averaged in time. The spatial resolution was 40~\si{\micro\meter}/\text{pixel}. Photoelastic measurements were performed at 20${}^\circ$C, as controlled by a temperature system on the rheometer. The calculations based on Eq. \eqref{eq:retardation} using the CRYSTA Stress Viewer software (Photron) were analyzed using ImageJ, MATLAB (R2023b), and Python.

The stress-optic coefficient $C_2$ can be obtained by considering the second-order SOL in rheo-optical measurements under Couette flow in a rheometer.
The viscous stress tensor in a rheometer $\mathbf{T_{rheo}}$ is as follows:
\begin{equation}
\mathbf{T_{rheo}} = 
\begin{pmatrix}
\textcolor{black}{0}
& 0
& -A y\\
0 
& \textcolor{black}{0}
& A x\\\
-A y
& A x
& \textcolor{black}{0}\\
\end{pmatrix},\label{eq:stress_tensor_rheo}
\end{equation}
where $A$ = $-2T/\pi R_{\mathrm{0}}^4$, $T$ is the torque measured by the rheometer, and $R_0$ is the plate radius (see Appendix \ref{appA}). These stress components are substituted into Eqs. \eqref{eq:v1} and \eqref{eq:v2}, from which we obtain the following relationship:
\begin{equation}
C_2=\frac{\Delta\textcolor{black}{(\dot{\gamma}(r))^k}}{d}\cdot \frac{\pi^2R_{\mathrm{0}}^6}{4T^2}\cdot \left(\frac{\dot{\gamma}_\mathrm{max}}{\dot{\gamma}(r)}\right)^2.
\label{eq:C2_calc}
\end{equation}
where $\dot{\gamma}$ is the shear rate and $\dot{\gamma}_\mathrm{max}$ = $\dot{\gamma}(R_o)$ is the maximal shear rate at $r = R_o$. According to the previous studies (\cite{lane2022birefringent,worby2024examination}),  $\Delta$ is proportional to  $\dot{\gamma}^{k}$, where $k$ denotes an effective local exponent characterizing the empirical scaling relationship $\Delta$ proportional to $\dot\gamma^k$, which has been reported in previous rheo-optical studies of birefringent suspensions (e.g., \cite{lane2022birefringent,worby2024examination}).

\begin{figure}
\centering
\includegraphics[width=1.0\linewidth]{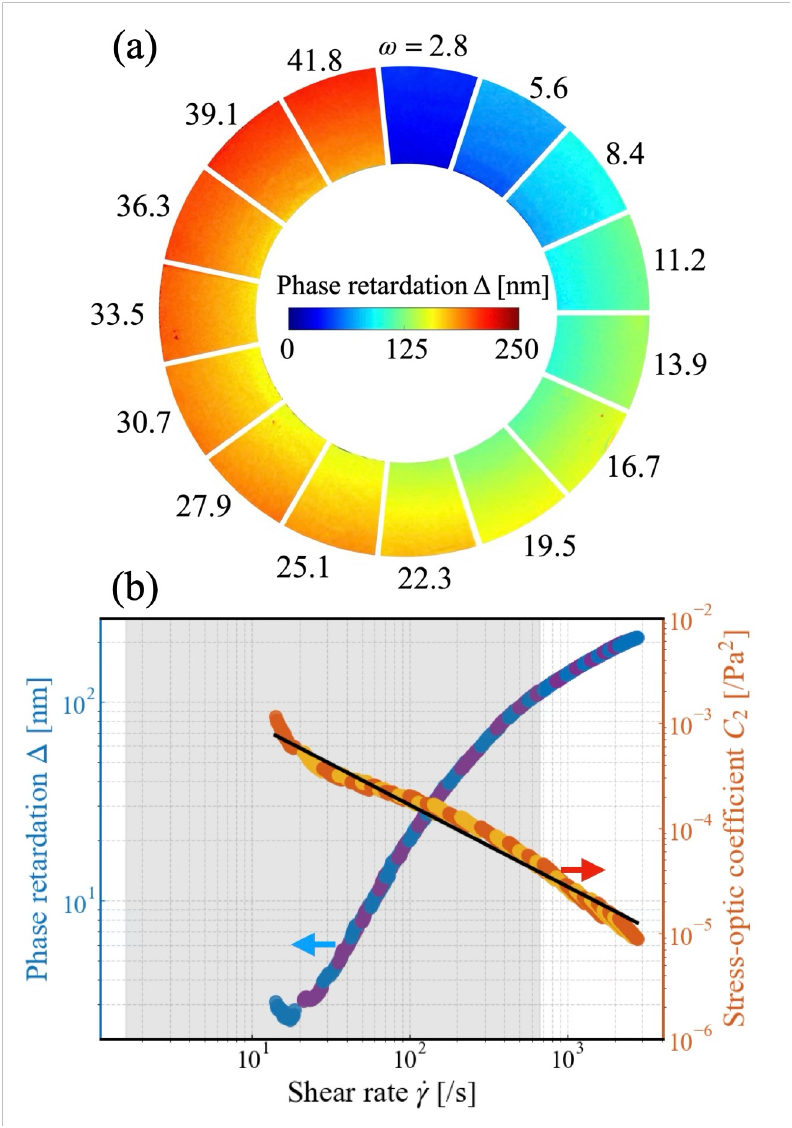}
\caption{(a) Phase retardation measured using rheo-optical setup (Fig. \ref{fig:rheometer}) for various shear rates. (b) Phase retardation $\Delta$ and $C_2$ as a function of shear rate $\dot{\gamma}$. Blue and purple plots are measured $\Delta$ and yellow and red plots are $C_2$ calculated using Eq.~\eqref{eq:C2_calc}. Different plot colors indicate results with different shear rate settings in the rheometer. Black solid line is the fitting curve given by Eq.~\eqref{eq:C2_fit}. \R{Region highlighted in gray corresponds to the shear rate range encountered in the Hele-Shaw experiments.}}
\label{fig:C2}
\end{figure}

The measured phase retardation using rheometer for various shear rates is shown in Fig. \ref{fig:C2}(a) and plotted in Fig. \ref{fig:C2}(b) as a function of shear rate. As the shear rate changes in the radial direction because of the parallel plates, the phase retardation for a certain range of shear rates can be obtained from a single measurement. The different colors in Fig. \ref{fig:C2}(b) indicate results with different shear rate settings. \R{In addition, the region highlighted in gray corresponds to the shear rate range encountered in the Hele-Shaw experiments. A small portion of the data falls below the calibrated shear-rate range due to the measurement limit of the rheometer. Nevertheless, across all conditions, less than 5\% of the domain in the $r$–$z$ plane exhibits shear rates below $10~\mathrm{s}^{-1}$; thus, the main trends and conclusions are governed by the shear-rate regime that is well covered by the calibration.}

According to \citet{worby2024examination}, the coefficient of the second term of the SOL ($C_2$) is a function of the second invariant of the deformation tensor even for Newtonian fluid. To consider the contribution of three-dimensional effects of the stress field, where the stress components act in the direction of light propagation, $C_2$ is assumed to be expressed as a function of the shear rate. Although those relationship was supported by experimental results, the reason why $C_2$ depends on shear rate has not yet been clarified. Fitting of the measured data gives the following relationship:
\begin{equation}
C_2 = \alpha |\dot{\gamma}|^\beta,
\label{eq:C2_fit}
\end{equation}
where $\alpha = 0.0059$ \R{(95\% confidence interval: $0.0056$--$0.0062$)},$~ \beta = -0.7759$ \R{(95\% confidence interval:$-0.7903$--$-0.7615$)} \R{with the coefficient of determination, $R^2=0.947$}. Since $\Delta$ is proportional to $\dot{\gamma}^{k}$, $k-2$ is represented as $\beta$. \citet{worby2024examination} reported that the fitting parameters were $\alpha = 4.4 \times 10^{-6}$ and $\beta = \textcolor{black}{-0.724}$, and stated that $\alpha$ is a function of the concentration of rod particles, indicating a birefringent material. The difference between the values of $\alpha$ and $\beta$ obtained in the present study and those reported by \citet{worby2024examination} is likely to be caused by differences in particle alignment due to the different sizes of rod particles, especially length.

Based on these results, the coefficient $C_2$ used in the theoretical calculation of the photoelastic retardation (Eqs. \eqref{eq:delta}--\eqref{eq:v2}) was given by Eq. \eqref{eq:C2_fit} so that $C_2$ was a function of the shear rate. 

From Fig. \ref{fig:C2}, the phase retardation increases with increasing shear rate. This is because the CNC particles oriented to the flow direction. As the orientation of the CNC particles reaches the flow direction, the phase retardation asymptotically approaches a constant value. In contrast, the stress-optic coefficient $C_2$ decreases with increasing shear rate.
Although both trends appear to follow a power-law relationship, they do not perfectly conform to a power-law.
For retardation, according to previous studies \citep{lane2022birefringent,calabrese2022alignment,worby2024examination}, there is a power-law relation between the phase retardation and the shear rate. However, the measured data do not perfectly follow a power-law relationship.
One possible reason for this deviation from the power-law relationship is the particle interactions that occur because of the relatively high concentration compared with previous studies.

\section{Results and Discussion}\label{r&d}

\begin{figure}
\centering
\includegraphics[width=1\linewidth]{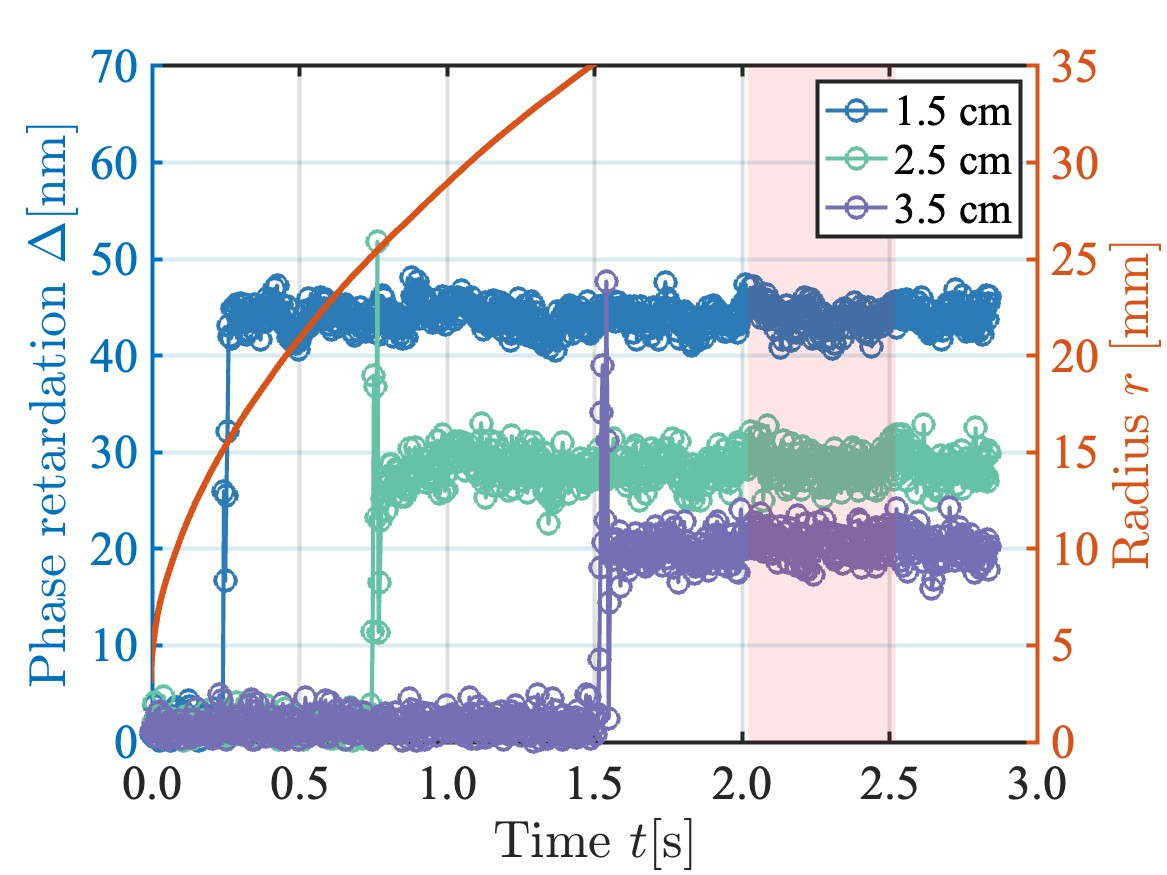}
\caption{Time history of phase retardation measured at fixed positions located 1.5 cm, 2.5 cm, and 3.5 cm in the direction of $\theta$ = 180$^{\circ}$ from the center of the inlet (in blue, green, purple; left $y$-axis) and measured radius of the flow field (in red; right $y$-axis) for $Q$ = 50 ml/min. Region highlighted in red corresponds to the section for which time averaging was applied.}
\label{fig:time_delta}
\end{figure}

To evaluate the steady state of phase retardation in radial Hele-Shaw flow, the time evolution of the retardation field was investigated. Figure \ref{fig:time_delta} shows the time history of phase retardation measured at three fixed positions ($r$ = 1.5 cm, 2.5 cm, and 3.5 cm in the direction of $\theta$ = 180$^{\circ}$) and the radius of the interface from the center of the inlet. From Fig. \ref{fig:time_delta}, phase retardation of approximately 5 nm occurs before the interface reaches each fixed measurement point, even in the absence of fluid. This can be interpreted as an optical measurement error. \R{Note that the baseline phase retardation (about 5 nm), observed in the absence of fluid, was not subtracted from the reported data. Since this offset is constant and small compared to the measured retardation (typically larger than 10 nm), it does not affect the overall trends or conclusions.}  After the interface passes through the fixed measurement points, the phase retardation increases sharply, and then oscillates around a constant value of 5 nm for all measurement points. Similar oscillation amplitudes occur under other conditions. These oscillations are caused by optical measurement error and machinery error associated with the syringe pump. Thus, the phase retardation was time averaged over a 0.5-s period starting 0.5 s after the averaged interface radius reached 35 mm (highlighted in red in Fig. \ref{fig:time_delta}).

\begin{figure*} 
\centering
\includegraphics[width=1\linewidth]{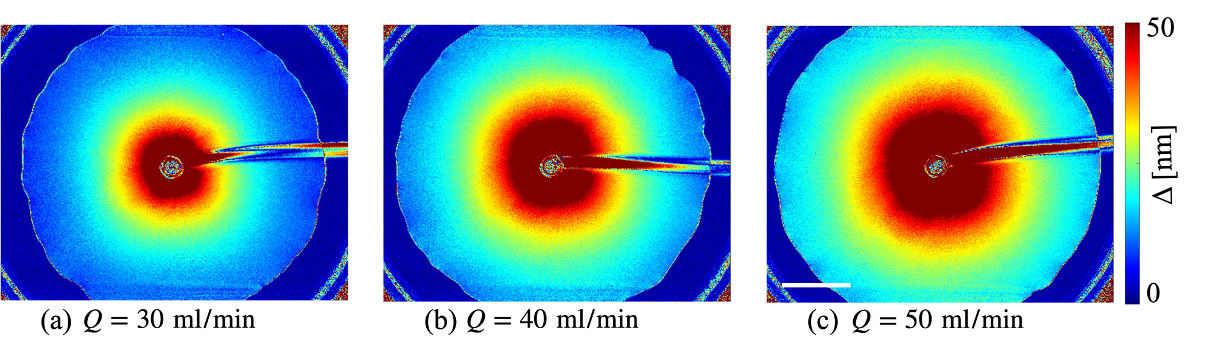}
\caption{Examples of measured phase retardation distribution in Hele-Shaw cell for various flow rates at the end of the temporal averaging period (highlighted in Fig. \ref{fig:time_delta}). White bar indicates scale for 20 mm. (a) $Q$ = 30 ml/min, $t$ = 3.64 s, (b) $Q$ = 40 ml/min, $t$ = 2.85 s, (c) $Q$ = 50 ml/min, $t$ = 2.52 s. \R{The same color scale is used for all panels.}}
\label{fig:delta_map}

\centering
\includegraphics[width=0.85\linewidth]{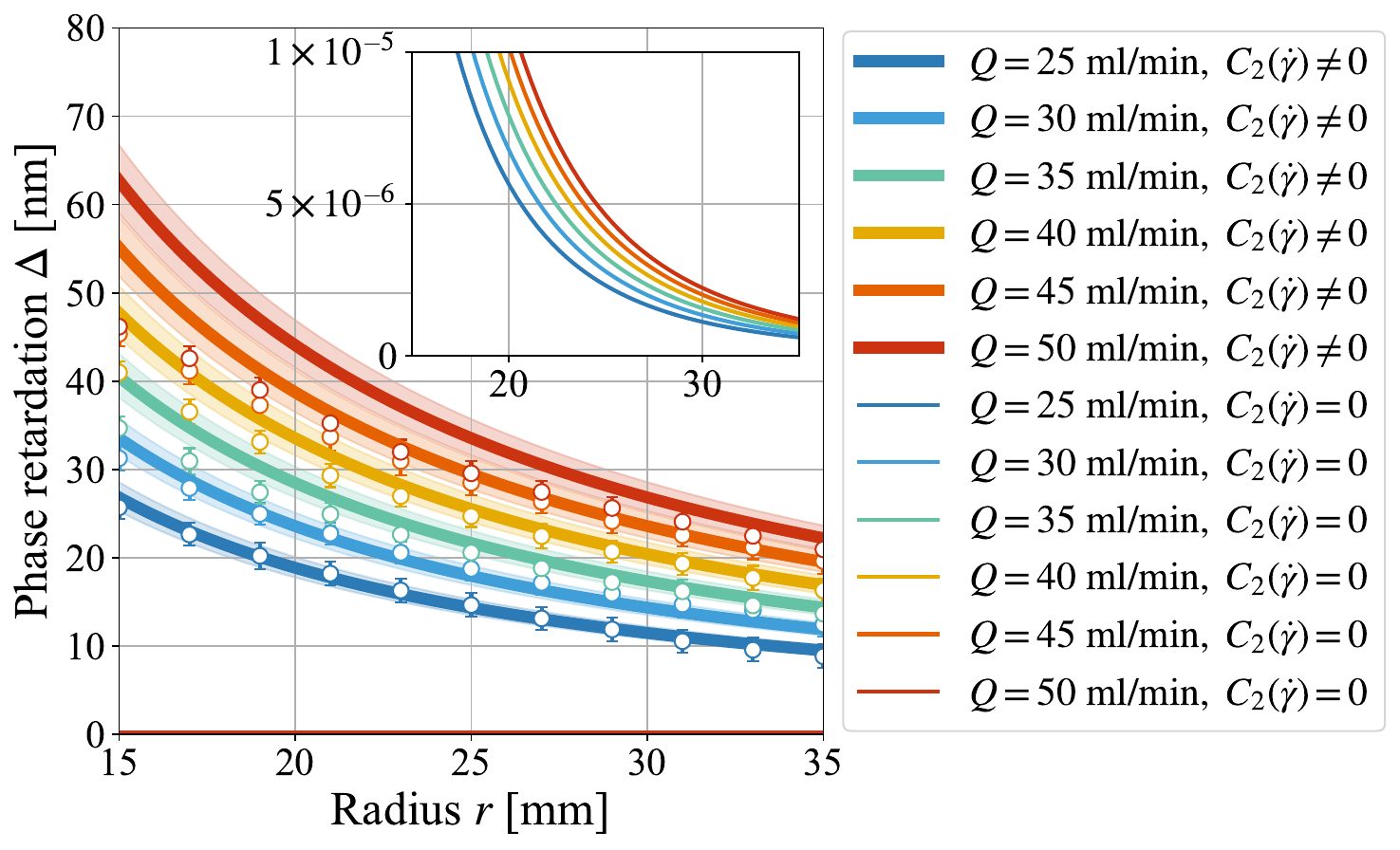}
\caption{Phase retardation in radial direction with theoretical results with a gap $b$ = 0.28\R{3} mm (solid lines) and experimental results (plots) for various flow rates $Q$ = 25--50 ml/min. Error bar \R{of plot} indicates $\pm$1SD \R{from three independent measurements}. \R{Shaded area represents the range $b$ = 0.283$\pm$0.012 mm. }Solid thick lines are theoretical solutions with $C_2(\dot{\gamma})(\neq0$ using Eq. \eqref{eq:C2_fit}) and solid thin lines are those with $C_2(\dot{\gamma}) = 0$. Inset highlights the region close to $\Delta$ = 0. \R{Note that the inset shows the same quantity on a different scale $(O(10^{-5}))$ for clarity, as the variations that are not visible in the main panel.}}
\label{fig:delta_comp}

\centering
\includegraphics[width=0.95\linewidth]{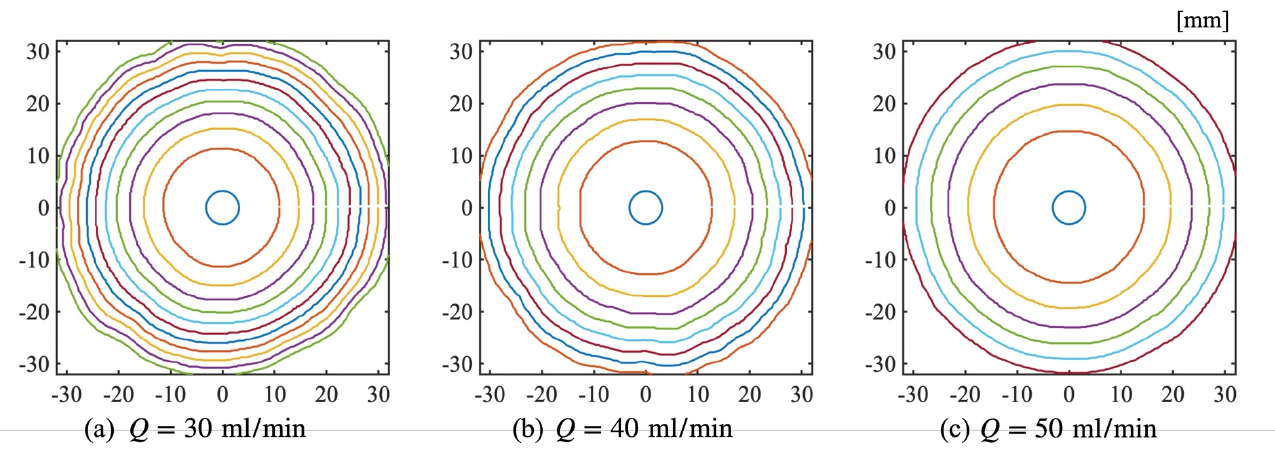}
\caption{Time evolutions of interface for various flow rates. The different colors correspond to a time interval of 0.2 s. (a) $Q$ = 30 ml/min, (b) $Q$ = 40 ml/min, (c) $Q$ = 50 ml/min.}
\label{fig:interface}
\end{figure*}

Figure \ref{fig:delta_map} depicts snapshots of the phase retardation distribution in the radial Hele-Shaw cell for various flow rates at the end of the averaging period. The phase retardation decreases in the radial direction for all cases and increases with increasing flow rate. When the interface shape deviates from a perfect circle, the phase retardation fluctuates in the $\theta$ direction.
From the observation shown in Fig. \ref{fig:delta_map}, since the variation in the $\theta$ direction is small and the phase retardation distribution is in accordance with the shear rate, the concentration in the $r$ and $\theta$ directions is expected to be almost homogeneous. The concentration distribution in the $z$-direction is likely to be influenced by inertial migration. The particle Reynolds number $Re_p$ is estimated at most order of $10^{-6}$, where $Re_\mathrm{p}$ = $Re\chi^2$, $Re = 2\rho Q/\pi\mu r$, and confinement $\chi$ is defined as $d_\mathrm{p}/b$ (The diameter $d_\mathrm{p}$ was defined as that of a sphere having the same volume as one CNC particle). Therefore, the inertial migration is negligible. Therefore, the particle distribution in the $z$ direction is also expected to be almost homogeneous.

Based on the results presented so far, the spatiotemporal averaged phase retardation is compared with the theoretical results. Figure \ref{fig:delta_comp} compares the experimental data (plots) with the theoretical solutions with $b$ = 0.28\R{3} mm (solid lines). \R{Error bar of plot indicates $\pm1$SD from three independent measurements. The precision of phase retardation was evaluated from repeated measurements, and the mean standard deviation ranged from 1.01 to 1.85 nm.}
\R{Since the slope of the $r-\Delta$ curve becomes steep near the inlet, the spatial resolution of 226 \textmu m/pixel (effective spatial resolution of approximately 450 \textmu m due to super-pixels) limits accuracy near the inlet (small $r$). Considering the limitation in spatial resolution, regions with steep gradients are excluded, and the region of interest is conservatively restricted to 15 mm $<r<$ 35 mm, where the spatial variations can be sufficiently resolved.}
The solid thick line indicates the theoretical solution with $C_2$ as a function of shear rate (Eq. \eqref{eq:C2_fit}), as estimated from the rheo-optical measurements (\S \ref{RO_measure}). The theoretical results reproduced the experimental data by taking $C_2$ into account. \R{Shaded area represents the range $b$ = 0.283$\pm$0.012 mm. By introducing a change of variables ($q= 2z/b$), the integral can be evaluated analytically as shown below. The second term scales with $1/b$, while the first term is independent of $b$.} 

\begin{equation}
\begin{aligned}
\Delta =  \frac{2C_1\mu Q}{\pi r^2}
-\frac{3C_2\mu^2 Q^2}{4\pi^2 r^2b}
\end{aligned}
\label{eq:delta_int}
\end{equation}

The solid thin line in the inset of Fig. \ref{fig:delta_comp} indicates the theoretical solution under the assumption of $C_2$ = 0, which corresponds to the solution with the conventional SOL. The theoretical results with $C_2$ = 0 are of $O(10^{-6})$. In Hele-Shaw flow, the stress components $\sigma_{rz}$ are larger than $\sigma_{rr}$ and $\sigma_{\theta \theta}$; therefore, the effects of $C_2$ are dominant over those of $C_1$. This suggests that the conventional SOL cannot capture the phase retardation in Hele-Shaw flow.

The difference between experimental and theoretical results increases with increasing flow rate. The difference is more pronounced near the inlet than in the outer region. This trend indicates that the slope of $\Delta-r$ curve in the experiment is more gradual compared with the theoretical solution for higher flow rates. This observed trend can be explained by two factors: the assumption of Newtonian fluid in the theoretical solution and the non-Newtonian (shear-thinning) property of the working fluid. As illustrated in Fig.~\ref{fig:shear_vis}, the viscosity in theoretical solution is underestimated in low-shear regions and overestimated in high-shear regions. Consequently, the theoretical retardation derived from Eq.~\eqref{eq:delta_int} is underestimated far from the inlet (in low-shear regions) and overestimated near the inlet (in high-shear regions). This relationship suggests that a proper consideration of shear-thinning behavior would result in a more gradual slope of the phase retardation in the radial direction. Therefore, one possible reason for the steeper slope of the theoretical solution compared with the experimental results is assumption of Newtonian fluid for working fluid.

The uncertainty associated with the averaging of viscosity was estimated to be approximately from -11\% to 36\% in fitted shear-rate range in 10<$\dot{\gamma}$<700, which corresponds to the interested conditions (25<$Q$<50 ml/min in 15<$r$<35 mm). \R{The accuracy was assessed by comparison with theoretical predictions, yielding a relative error from -25.9 to 2.6\% in 15 mm < $r$ < 35 mm.} Therefore, the approximation may contribute to the uncertainty in the results. Investigating the influence of non-Newtonian properties on flow birefringence remains a topic for future work.

The theoretical results show a smooth exponential decay as a function of $r$. In contrast, at lower flow rates, especially $Q = 25, 30$ ml/min, the experimental results slightly deviate from this monotonic behavior. One possible reason of this result is the interface shape.

The time evolution of the interface for various flow rates is shown in Fig. \ref{fig:interface}. In the initial phase with a small radius and high interfacial velocity, the test fluid spreads out in concentric rings, regardless of flow rate. As the radius increases and the interface velocity decreases, the interface deviates from the circle, and the deviation becomes more significant for lower flow rates.

\begin{figure}
\centering
\includegraphics[width=1\linewidth]{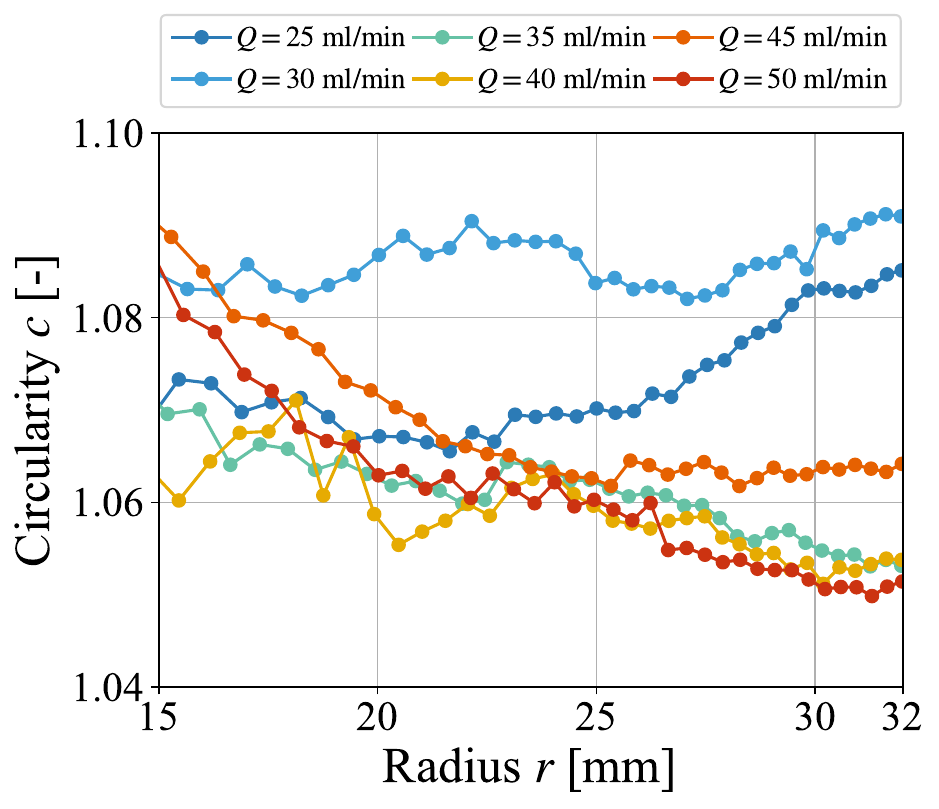}
\caption{\R{Circularity of interface shape for various flow rates.}}
\label{fig:circularity}
\end{figure}

To evaluate the interface shape quantitatively, the circularity was evaluated using the distance from the center of the inlet to the interface over 360$^{\circ}$. The circularity is defined as
\begin{equation}
c = \frac{r_\mathrm{max}}{r_\mathrm{min}},
\label{eq:circ}
\end{equation}
where $r_\mathrm{max}$ is the radius at the furthest interface from the center of inlet and $r_\mathrm{min}$ is the radius at the nearest interface from the center of inlet (Fig. \ref{fig:setup}). The circularity $c = 1$ denotes perfect circle and values greater than 1 ($c > 1$) indicate deviation from a circle. The relationship between the interface radius and circularity is shown in Fig. \ref{fig:circularity}. 
The interface shape for low flow rates deviates from a perfect circle, and becomes close to a circle with increasing flow rate. When the pressure gradient decreases with a large radius or low flow rate, the relative magnitude between the pressure gradient and surface tension changes locally. The pressure then increases due to local interface deformation, resulting in deviation from a circular shape. Although the experimental system is different, a similar observation was reported by \citet{suzuki2019fingering}, where they state that the interface becomes stable at high flow rates when the pressure gradient exceeds the thermodynamic body force.
When the interface shape deviates from a perfect circle, the phase retardation change in circumferential direction as shown in Fig. \ref{fig:delta_map}. Since the phase retardation itself is small for lower flow rates, the phase retardation change become relatively larger than that for high flow rate cases. 
The difference in the trend between low shear (large radius or low flow rate) region and high shear (near inlet or high flow rate) conditions is attributed to interfacial deformation.

In summary, the experimental phase retardation in a radial Hele-Shaw cell has been explained using the theoretical solution based on the second-order SOL. $C_2$ is considered as a function of shear rate based on rheo-optical measurements. To the best of our knowledge, this represents one of the first quantitative experimental studies of flow birefringence in Hele-Shaw flow.

An explanation of phase retardation in a Hele-Shaw cell is an important step toward our understanding of three-dimensional stress fields.
Hele-Shaw cells are often used to study the Saffman--Taylor instability (viscous fingering) \citep{saffman1958penetration,homsy1987viscous}. Flow birefringence has been used to study other instabilities, such as the elastic instability \citep{robert2003flow} and the stick-slip instability \citep{zhao2016flow}. Previous theoretical, experimental, and numerical approaches have found that the three-dimensional structure inside the gap plays an important role in the onset and suppression of viscous fingering \citep{lajeunesse19973d, yang1997asymptotic, lajeunesse1999miscible,bischofberger2014fingering,oliveira2011miscible,videbaek2019diffusion,nand2022effect}. 
Hele-Shaw cells are not only used to observe the Saffmann--Taylor instability, but are also important in studies on multiphase flows \citep{marin2015three, zhang2023three}. These studies illustrate the importance of the three-dimensional flow field in Hele-Shaw flow.
In this context, the results presented in this paper contribute to advances in the study of fluid mechanics.

\section{Conclusion}\label{conc}
We have measured the flow birefringence at various flow rates with CNC suspensions in a steady radial Hele-Shaw flow. The flow birefringence has been investigated, focusing on the effects of stress along the optical direction. The measured phase retardation was compared with the theoretical solution based on the SOL. We found that the conventional SOL does not correctly predict the experimental phase retardation in Hele-Shaw flow, as it neglects the stress along the optical axis. Phase retardation in Hele-Shaw flow, which cannot be described by the conventional SOL, was described using the second-order SOL based on rheo-optical measurements.
The results of our study will enable a quantitative interpretation of the measured flow birefringence in Hele-Shaw flow.
\R{The present measurements are limited to steady conditions, as time-resolved measurements of rapidly evolving flows remain challenging. Addressing non-Newtonian viscosity more rigorously will be an important direction for future work, together with the theoretical clarification of microscopic origin of $C_2(\dot{\gamma})$. The present framework may be also extended to interfacial instability problems such as viscous fingering.}

\section*{Acknowledgments}
R. X. S. and Y.T. thank the JST PRESTO (Grant No.~JPMJPR22O5), and M.K., R. X. S., Y.N. and Y.T. thank the Sumitomo Foundation (Grant No.~2401115) for financial support; M.K. acknowledges financial support from JSPS KAKENHI (Grant No.~JP23K13252), and Y.T. acknowledges financial support from JSPS KAKENHI (Grant No.~JP20H00223) and JST PRESTO (Grant No.~JPMJPR21O5).

\appendix

\setcounter{section}{0}
\renewcommand{\thesection}{\Alph{section}}
\renewcommand{\thefigure}{\Alph{section}.\arabic{figure}}
\renewcommand{\thetable}{\Alph{section}.\arabic{table}}
\setcounter{figure}{0}
\setcounter{table}{0}

\section*{Appendix}
\begin{appendix}

\section{Determination of $C_2$ from rheo-optical measurements} \label{appA}

The stress tensor of an incompressible fluid in cylindrical coordinates is \citep{bird2002}

\begin{equation}
\begin{aligned}
\mathbf{T}_{r\theta z} &
=
\scalebox{0.8}{$
\begin{pmatrix}
2\mu\frac{\partial u_r}{\partial r}&\mu \left( r\frac{\partial }{\partial r}\left( \frac{u_r}{r} \right) + \frac{1}{r}\frac{\partial u_r}{\partial \theta} \right)& \mu\left(\frac{\partial u_r}{\partial z} + \frac{\partial u_z}{\partial r} \right)\\[10pt]
\mu \left( r\frac{\partial }{\partial r}\left( \frac{u_r}{r} \right) + \frac{1}{r}\frac{\partial u_r}{\partial \theta} \right)& 2\mu\left(\frac{1}{r}\frac{\partial u_\theta}{\partial \theta}+\frac{u_r}{r}\right) & \mu \left( \frac{1}{r}\frac{\partial u_z}{\partial \theta}+\frac{\partial u_\theta}{\partial z}\right) \\[10pt]
\mu\left(\frac{\partial u_r}{\partial z} + \frac{\partial u_z}{\partial r} \right)& \mu \left( \frac{1}{r}\frac{\partial u_z}{\partial \theta}+\frac{\partial u_\theta}{\partial z}\right) & 2\mu\frac{\partial u_z}{\partial z}
\end{pmatrix}.
$}
\end{aligned}
\label{eq:stress_rtz}
\end{equation}

In the flow within a rheometer using a parallel plate, $\frac{\partial}{\partial \theta}$ is zero under the assumption of axial symmetry with respect to the $z$-axis. In addition, the velocities $u_\theta$ and $u_z$ are assumed to be zero. Then,

\begin{equation}
\begin{aligned}
\mathbf{T}_{r\theta z} 
&=
\begin{pmatrix}
\textcolor{black}{0} & 0 & 0 \\[10pt]
0 & \textcolor{black}{0} & \mu\frac{\partial u_\theta}{\partial z} \\[10pt]
0 & \mu\frac{\partial u_\theta}{\partial z} & \textcolor{black}{0}
\end{pmatrix}
=\begin{pmatrix}
\textcolor{black}{0} & 0 & 0 \\[10pt]
0 & \textcolor{black}{0} & \frac{2rT}{\pi \mu R_o^4} \\[10pt]
0 & \frac{2rT}{\pi \mu R_o^4} & \textcolor{black}{0}
\end{pmatrix},\\
\end{aligned}
\end{equation}
where the velocity $u_\theta$ in a viscometer with a parallel plate \citep[pp.37-38]{bird1987dynamics} is

\begin{equation}
\begin{aligned}
u_\theta = \frac{r\omega}{d}z = \frac{2rT}{\pi \mu R_o^4}z.
\end{aligned}
\end{equation}

Transforming from cylindrical coordinates to Cartesian coordinates, the stress tensor is expressed as
\begin{equation}
\begin{aligned}
\mathbf{T}_{xyz}=\begin{pmatrix}
\textcolor{black}{0} & 0 & \frac{2T}{\pi \mu R_o^4}y \\[10pt]
0 & \textcolor{black}{0} & -\frac{2T}{\pi \mu R_o^4}x \\[10pt]
\frac{2T}{\pi \mu R_o^4}y & -\frac{2rT}{\pi \mu R_o^4}x & \textcolor{black}{0}
\end{pmatrix}.\\
\end{aligned}
\end{equation}
Finally, we define $\mathbf{T}_{xyz}=\mathbf{T}_\text{rheo}$.

\section{Effects of $C_1$ on theoretical phase retardation} \label{appB}

\begin{figure}
\centering
\includegraphics[width=1\linewidth]{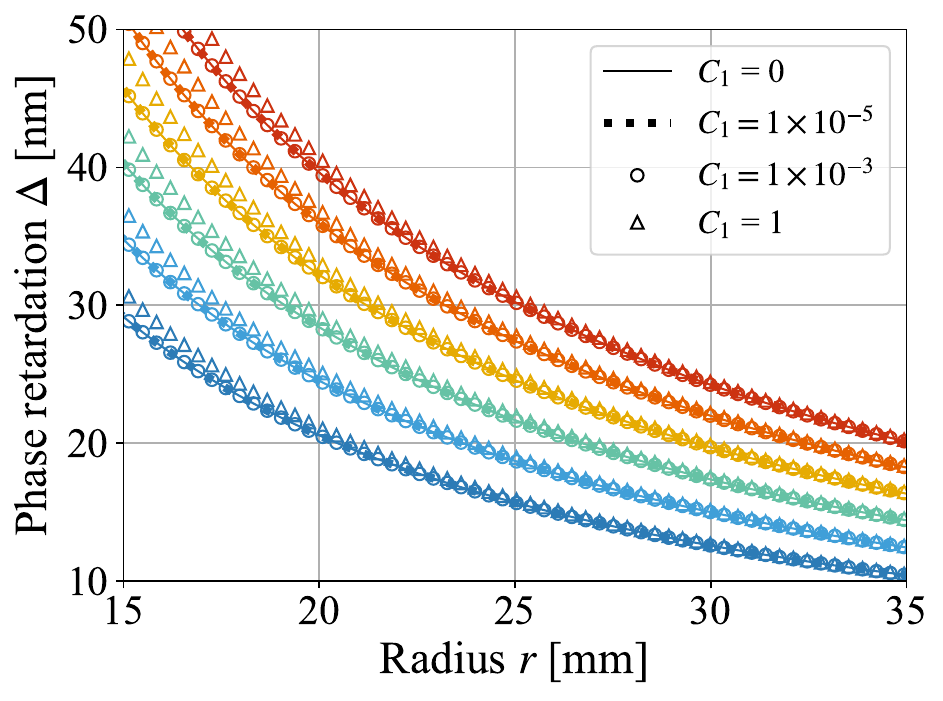}
\caption{Theoretical phase retardation in radial direction for various $C_1$. Color corresponds to the different flow rate 25 ml/min $< Q <$ 50 ml/min same as Fig.\ref{fig:delta_comp}.}
\label{fig:c1}
\end{figure}

Fig.~\ref{fig:c1} shows the theoretical solution with cell gap of $b = 0.28$ mm for various $C_1$. \cite{nakamine2024flow} reported that the $C_1$ is linearly proportional to the CNC concentration . Based on this data, $C_1$ is approximately $C_1 \sim O(10^{-5})$ for 3 wt\% CNC suspension. The variation of $C_1$ up to $10^{-3}$ has little influence on the phase retardation in Hele-Shaw configuration.

\section{\R{Effects of shear-thinning on theoretical phase retardation}} \label{app_non-newtonian}
\R{To examine the effects of the shear-thinning, the shear viscosity is fitted using the following power-law model.}

\begin{align}
\mu_\text{p}(\dot{\gamma}) = m\dot{\gamma}^{n-1} ,
\label{eq:power_vis}
\end{align}
\R{where $m$ and $n$ are power-law constants, and $\dot{\gamma}$ is the shear rate. }
\R{The power-law coefficients $m$ = 8.7429 and $n$ = 0.8951 are used in Eq. \eqref{eq:power_vis} based on the fitting of measured data with sonication for 400 s.}

\R{The velocity profile for the non-Newtonian case can be expressed using a power-law solution \citep{macosko1994rheology,bird1987dynamics} as}

\begin{align}
u_r(r,z) &= 
\frac{Q}{2\pi rb} \frac{2n+1}{n+1}
\left(
1-\left|\frac{2z}{b}\right|^{\frac{n+1}{n}}
\right) ,
\label{eq:GNF_vel}
\end{align}
\R{where $Q$ is the flow rate. Equation \eqref{eq:GNF_vel} is taken from \citet{winter1975approximate}, with the modification of the absolute value in the last term. The absolute value is applied to prevent negative values inside the exponent, ensuring a real-valued result.} 
\R{The shear-thinning effects are considered in the velocity (Eq. \eqref{eq:GNF_vel}) and viscosity in the stress tensor (Eq. \eqref{eq:stress_tensor}). The theoretical solution of phase retardation with the shear-thinning effect is calculated as} 
\begin{equation}
\begin{aligned}
\Delta &= \sqrt{V_1^2+V_2^2} = \left|V_1\right| \\
&= \int_{-b/2}^{b/2} \bigl(2C_1{\sigma_{\theta\theta}} - C_2(\dot{\gamma}){\sigma_{rz}}^2\bigr)dz.\\
&= \int_{-b/2}^{b/2} \Bigg[\frac{2C_1\mu_\text{p}(\dot{\gamma}) Q}{\pi r^2b}\frac{2n+1}{n+1}
\left(1-\left|\frac{2z}{b} \right|^{\frac{n+1}{n}}\right)\\
&\quad
-C_2(\dot{\gamma})\left(\frac{\mu_\text{p}(\dot{\gamma}) Q}{2\pi rb}\frac{2n+1}{n}
\left|\frac{2z}{b} \right|^{\frac{1}{n}} \right)^2 \Bigg]dz.
\end{aligned}
\label{eq:delta_analytical_nn}
\end{equation}

\begin{figure}
\centering
\includegraphics[width=1\linewidth]{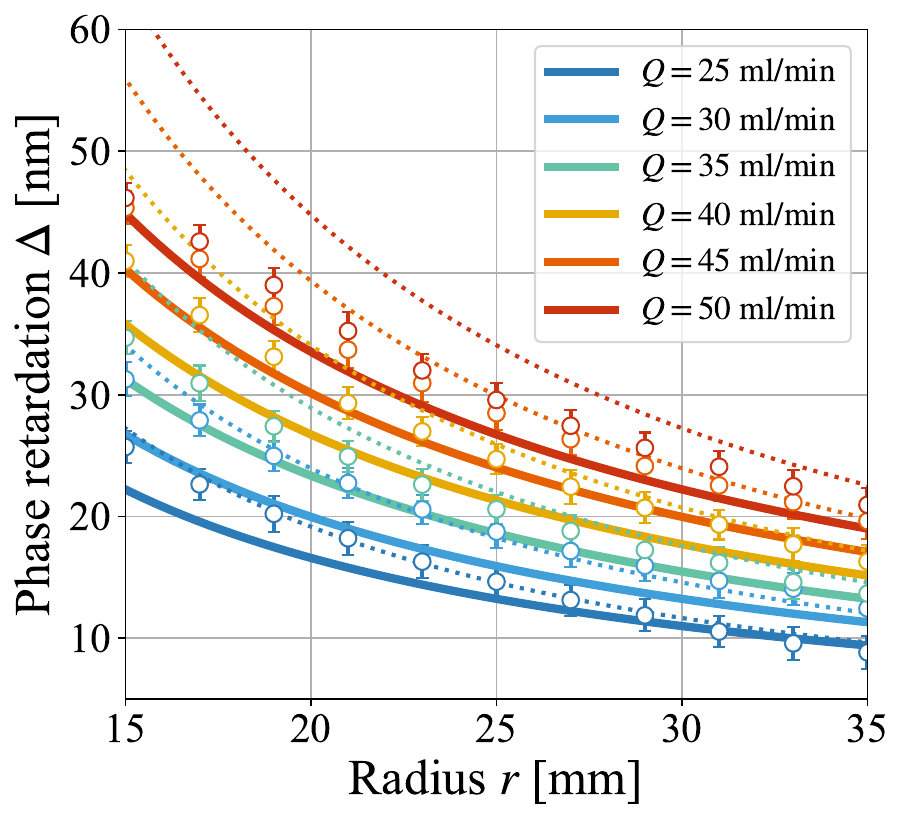}
\caption{\R{Phase retardation distribution in radial direction for $b$ = 0.283 mm for flow rates $Q$ = 25 -- 50 ml/min. The solid line assumes $\mu$ is constant and the dotted line assumes shear-thinning viscosity is a function of shear rate fitted by the power-law model (Eq. \eqref{eq:power_vis}). Plots are experimental results and error bar indicates $\pm$1SD from three independent measurements. }}
\label{fig:delta-r_NN}
\end{figure}

\R{Figure \ref{fig:delta-r_NN} shows a comparison of theoretical phase retardation obtained under Newtonian and non-Newtonian assumptions. As a result, the simple power-law approximation confirms that shear-thinning can account for the observed reduction in slope mismatch. This supports the interpretation that the present comparison provides indicative evidence of the limitation of the Newtonian assumption.}

\R{This comparison is intended as a supplementary assessment of shear-thinning effects, while the theoretical prediction used for comparison with the experimental data is based on a Newtonian assumption. A more rigorous treatment of non-Newtonian effects is beyond the scope of the present study.}

\end{appendix}










\printcredits

\bibliographystyle{cas-model2-names}

\bibliography{cas-refs}

@book{doi1988theory,
  title={The theory of polymer dynamics},
  author={Doi, Masao and Edwards, Samuel Frederick},
  volume={73},
  year={1988},
  publisher={oxford university press}
}

@book{macosko1994rheology,
title={Rheology principles},
author={Macosko, Christopher W},
journal={Measurements and Applications},
year={1994},
publisher={VCH Publishers}
}

@book{bird1987dynamics,
title={Dynamics of polymeric liquids. Vol. 1: Fluid mechanics},
author={Bird, Robert Byron and Armstrong, Robert Calvin and Hassager, Ole},
year={1987},
publisher={John Wiley and Sons Inc., New York, NY}
}

@article{winter1975approximate,
title={Approximate calculation and measurement of the pressure distribution in radial flow of molten polymers between parallel discs},
author={Winter, Horst H},
journal={Polym Eng Sci},
volume={15},
number={6},
pages={460--469},
year={1975},
publisher={Wiley Online Library}
}

@book{bird2002,
author = "Bird, R.B. and Stewart, W.E. and Lightfoot, E.N.",
title = "Transport Phenomena, 2nd ed.",
address = "New York, NY, USA",
publisher = "Wiley Text Books",
year = "2002" 
}

@article{decruppe2003flow,
title={Flow birefringence, stress optical rule and rheology of four micellar solutions with the same low shear viscosity},
author={Decruppe, JP and Ponton, A},
journal={Eur Phys J E},
volume={10},
pages={201--207},
year={2003},
publisher={Springer}
}

@article{prabhakaran1975stress,
title={On the stress-optic law for orthotropic-model materials in biaxial-stress fields },
author={Prabhakaran, R},
journal={Exp Mech},
volume={15},
pages={29--34},
year={1975},
publisher={Springer}
}

@book{batchelor2000introduction,
title={An introduction to fluid dynamics},
author={Batchelor, George Keith},
year={2000},
publisher={Cambridge University Press}
}

@article{deshmukh2022cleaning,
title={Cleaning of simple cohesive soil layers in a radial flow cell},
author={Deshmukh, KP and Arlov, D and Cant, RS and G{\"o}ransson, A and Innings, F and Wilson, DI},
journal={Food Bioprod Process},
volume={136},
pages={84--96},
year={2022},
publisher={Elsevier}
}

@article{delon2020hele,
title={Hele-{S}haw microfluidic device: A new tool for systematic investigation into the effect of the fluid shear stress for organs-on-chips},
author={Delon, Ludivine C and Guo, Zhaobin and Kashani, Moein Navvab and Yang, Chih-Tsung and Prestidge, Clive and Thierry, Benjamin},
journal={MethodsX},
volume={7},
pages={100980},
year={2020},
publisher={Elsevier}
}

@article{zhang2023three,
title={Three-dimensional streaming around an obstacle in a {H}ele-{S}haw cell},
author={Zhang, Xirui and Rallabandi, Bhargav},
journal={J Fluid Mech},
volume={961},
pages={A35},
year={2023}
}

@article{marin2015three,
title={Three-dimensional phenomena in microbubble acoustic streaming},
author={Marin, Alvaro and Rossi, Massimiliano and Rallabandi, Bhargav and Wang, Cheng and Hilgenfeldt, Sascha and K{\"a}hler, Christian J},
journal={Phys Rev Appl},
volume={3},
number={4},
pages={041001},
year={2015},
publisher={APS}
}

@article{homsy1987viscous,
title={Viscous fingering in porous media},
author={Homsy, George M},
journal={Annu Rev Fluid Mech},
volume={19},
number={1},
pages={271--311},
year={1987},
publisher={Annual Reviews 4139 El Camino Way, PO Box 10139, Palo Alto, CA 94303-0139, USA}
}

@book{ramesh2021developments,
title={Developments in Photoelasticity: A renaissance},
author={Ramesh, Krishnamurthi},
year={2021},
publisher={IOP Publishing}
}

@article{onuma2014development,
title={A development of two-dimensional birefringence distribution measurement system with a sampling rate of 1.3 {MHz}},
author={Onuma, Takashi and Otani, Yukitoshi},
journal={Opt Commun},
volume={315},
pages={69--73},
year={2014},
publisher={Elsevier}
}

@article{otani1994two,
title={Two-dimensional birefringence measurement using the phase shifting technique},
author={Otani, Yukitoshi and Shimada, Takuya and Yoshizawa, Toru and Umeda, Norihiro},
journal={Opt Eng},
volume={33},
number={5},
pages={1604--1609},
year={1994},
publisher={SPIE}
}

@article{suzuki2019fingering,
title={Fingering pattern induced by spinodal decomposition in hydrodynamically stable displacement in a partially miscible system},
author={Suzuki, Ryuta X and Nagatsu, Yuichiro and Mishra, Manoranjan and Ban, Takahiko},
journal={Phys Rev Fluids},
volume={4},
number={10},
pages={104005},
year={2019},
publisher={APS}
}

@article{lane2022birefringent,
title={Birefringent properties of aqueous cellulose nanocrystal suspensions},
author={Lane, Connor and Rode, David and R{\"o}sgen, Thomas},
journal={Cellulose},
volume={29},
number={11},
pages={6093--6107},
year={2022},
publisher={Springer}
}

@article{sato2024two,
title={Two-dimensional rheo-optical measurement system to study dynamics and structure of complex fluids},
author={Sato, Taisuke and Yamagata, Yoshifumi and Sato, Yasunori and Onuma, Takashi and Miyamoto, Keisuke and Takahashi, Tsutomu},
journal={Appl Rheol},
volume={34},
number={1},
pages={20240006},
year={2024},
publisher={De Gruyter}
}

@article{zhao2016flow,
title={Flow of wormlike micellar solutions around confined microfluidic cylinders},
author={Zhao, Ya and Shen, Amy Q and Haward, Simon J},
journal={Soft Matter},
volume={12},
number={42},
pages={8666--8681},
year={2016},
publisher={Royal Society of Chemistry}
}

@article{ober2011spatially,
title={Spatially resolved quantitative rheo-optics of complex fluids in a microfluidic device},
author={Ober, Thomas J and Soulages, Johannes and McKinley, Gareth H},
journal={J Rheol},
volume={55},
number={5},
pages={1127--1159},
year={2011},
publisher={AIP Publishing}
}

@article{oliveira2011miscible,
title={Miscible displacements in {H}ele-{S}haw cells: three-dimensional {N}avier--{S}tokes simulations},
author={Oliveira, Rafael M and Meiburg, Eckart},
journal={J Fluid Mech},
volume={687},
pages={431--460},
year={2011},
publisher={Cambridge University Press}
}

@article{kloosterman2011flow,
title={Flow rate estimation in large depth-of-field micro-{PIV}},
author={Kloosterman, A and Poelma, C and Westerweel, J},
journal={Exp Fluids},
volume={50},
pages={1587--1599},
year={2011},
publisher={Springer}
}

@article{ehyaei2014quantitative,
title={Quantitative velocity measurement in thin-gap {P}oiseuille flows},
author={Ehyaei, Dana and Kiger, Kenneth T},
journal={Exp Fluids},
volume={55},
pages={1--12},
year={2014},
publisher={Springer}
}

@article{sun2016measurements,
title={Measurements of flow-induced birefringence in microfluidics},
author={Sun, Chen-Li and Huang, Hung-Yen},
journal={Biomicrofluidics},
volume={10},
number={1},
year={2016},
pages={011903},
publisher={AIP Publishing}
}

@article{maxwell1873double,
title={On double refraction in a viscous fluid in motion},
author={Maxwell, J Clerk},
journal={Proc R Soc Lond},
volume={22},
pages={46--47},
year={1874},
publisher={JSTOR}
}

@article{sun1999visualisation,
title={Visualisation of dynamic flow birefringence of cardiovascular models},
author={Sun, Yong-Da and Sun, Yong-Fang and Sun, Yang and Xu, XY and Collins, MW},
journal={Opt Laser Technol},
volume={31},
number={1},
pages={103--112},
year={1999},
publisher={Elsevier}
}

@article{cerf1952flow,
title={Flow birefringence in solutions of macromolecules.},
author={Cerf, Roger and Scheraga, Harold A},
journal={Chem Rev},
volume={51},
number={2},
pages={185--261},
year={1952},
publisher={ACS Publications}
}

@article{hu2009flow,
title={Flow visualization using tobacco mosaic virus},
author={Hu, David L and Goreau, Thomas J and Bush, John WM},
journal={Exp Fluids},
volume={46},
pages={477--484},
year={2009},
publisher={Springer}
}

@article{rankin1989streaming,
title={A streaming birefringence study of the flow at the junction of the aorta and the renal arteries},
author={Rankin, GW and Sabbah, HN and Stein, PD},
journal={Exp Fluids},
volume={7},
number={2},
pages={73--80},
year={1989},
publisher={Springer}
}

@article{pindera1978characteristic,
title={Characteristic relations of flow birefringence: Part 1: Relations in transmitted radiation },
author={Pindera, JT and Krishnamurthy, AR},
journal={Exp Mech},
volume={18},
pages={1--10},
year={1978},
publisher={Springer}
}

@article{calabrese2022alignment,
title={Alignment of colloidal rods in crowded environments},
author={Calabrese, Vincenzo and Varchanis, Stylianos and Haward, Simon J and Shen, Amy Q},
journal={Macromolecules},
volume={55},
number={13},
pages={5610--5620},
year={2022},
publisher={ACS Publications}
}

@article{noto2020applicability,
title={Applicability evaluation of the stress-optic law in {N}ewtonian fluids toward stress field measurements},
author={Noto, Daisuke and Tasaka, Yuji and Hitomi, Jumpei and Murai, Yuichi},
journal={Phys Rev Res},
volume={2},
number={4},
pages={043111},
year={2020},
publisher={APS}
}

@article{doyle1982nonlinearity,
title={On a nonlinearity in flow birefringence },
author={Doyle, James F},
journal={Exp Mech},
volume={22},
pages={37--38},
year={1982},
publisher={Springer}
}

@book{aben2012photoelasticity,
title={Photoelasticity of glass},
author={Aben, Hillar and Guillemet, Claude},
year={1993},
publisher={Springer Science \& Business Media}
}

@article{lane2024two,
title={Two-dimensional strain rate imaging study using a polarization camera and birefringent aqueous cellulose nanocrystal suspensions},
author={Lane, Connor and Baumann, Fr{\'e}d{\'e}ric and Rode, David and R{\"o}sgen, Thomas},
journal={Exp Fluids},
volume={65},
number={1},
pages={8},
year={2024},
publisher={Springer}
}

@article{robert2003flow,
title={Flow birefringence study of the stick--slip instability during extrusion of high-density polyethylenes},
author={Robert, Laurent and Vergnes, Bruno and Demay, Yves},
journal={J Non-Newton Fluid Mech},
volume={112},
number={1},
pages={27--42},
year={2003},
publisher={Elsevier}
}

@article{nakamine2024flow,
title={Flow birefringence of cellulose nanocrystal suspensions in three-dimensional flow fields: Revisiting the stress-optic law},
author={Nakamine, Kento and Yokoyama, Yuto and Worby, William Kai Alexander and Muto, Masakazu and Tagawa, Yoshiyuki},
journal={Cellulose},
volume={31},
number={12},
pages={7405--7420},
year={2024},
publisher={Springer}
}

@article{worby2024examination,
title={Examination of flow birefringence induced by the shear components along the optical axis using a parallel-plate-type rheometer},
author={Worby, William Kai Alexander and Nakamine, Kento and Yokoyama, Yuto and Muto, Masakazu and Tagawa, Yoshiyuki},
journal={Sci Rep},
volume={14},
number={1},
pages={21931},
year={2024},
publisher={Nature Publishing Group UK London}
}

@article{hele1898flow,
title={Flow of water},
author={Hele-Shaw, Henry Selby},
journal={Nature},
volume={58},
number={1509},
pages={520--520},
year={1898},
publisher={Nature Publishing Group UK London}
}

@article{saffman1958penetration,
title={The penetration of a fluid into a porous medium or {H}ele-{S}haw cell containing a more viscous liquid},
author={Saffman, Philip Geoffrey and Taylor, Geoffrey Ingram},
journal={Proc R Soc Lond. Ser A. Math Phys Sci},
volume={245},
number={1242},
pages={312--329},
year={1958},
publisher={The Royal Society London}
}

@article{yang1997asymptotic,
title={Asymptotic solutions of miscible displacements in geometries of large aspect ratio},
author={Yang, Zhengming and Yortsos, Yanis C},
journal={Phys Fluids},
volume={9},
number={2},
pages={286--298},
year={1997},
publisher={American Institute of Physics}
}

@article{lajeunesse19973d,
title={{3D} instability of miscible displacements in a {H}ele-{S}haw cell},
author={Lajeunesse, Eric and Martin, Jerome and Rakotomalala, Nicole and Salin, Dominique},
journal={Phys Rev Lett},
volume={79},
number={26},
pages={5254},
year={1997},
publisher={APS}
}

@article{lajeunesse1999miscible,
title={Miscible displacement in a Hele-Shaw cell at high rates},
author={Lajeunesse, E and Martin, J and Rakotomalala, N and Salin, D and Yortsos, YC},
journal={Journal of Fluid Mechanics},
volume={398},
pages={299--319},
year={1999},
publisher={Cambridge University Press}
}

@article{bischofberger2014fingering,
title={Fingering versus stability in the limit of zero interfacial tension},
author={Bischofberger, Irmgard and Ramachandran, Radha and Nagel, Sidney R},
journal={Nat Commun},
volume={5},
number={1},
pages={5265},
year={2014},
publisher={Nature Publishing Group UK London}
}

@article{videbaek2019diffusion,
title={Diffusion-driven transition between two regimes of viscous fingering},
author={Videb{\ae}k, Thomas E and Nagel, Sidney R},
journal={Phys Rev Fluids},
volume={4},
number={3},
pages={033902},
year={2019},
publisher={APS}
}

@article{nand2022effect,
title={Effect of {H}ele-{S}haw cell gap on radial viscous fingering},
author={Nand, Sada and Sharma, Vandita and Das, Santanu Kumar and Padhee, Srikant Sekhar and Mishra, Manoranjan},
journal={Sci Rep},
volume={12},
number={1},
pages={18967},
year={2022},
publisher={Nature Publishing Group UK London}
}

@article{shafiei2012rheology,
title={Rheology of nanocrystalline cellulose aqueous suspensions},
author={Shafiei-Sabet, Sadaf and Hamad, Wadood Y and Hatzikiriakos, Savvas G},
journal={Langmuir},
volume={28},
number={49},
pages={17124--17133},
year={2012},
publisher={ACS Publications}
}

@article{rezania1997probabilistic,
title={A probabilistic approach to measure the strength of bone cell adhesion to chemically modified surfaces},
author={Rezania, Alireza and Thomas, Carson H and Healy, Kevin E},
journal={Ann Biomed Eng},
volume={25},
pages={190--203},
year={1997},
publisher={Springer}
}

@article{goldstein1998comparison,
title={Comparison of converging and diverging radial flow for measuring cell adhesion},
author={Goldstein, Aaron S and DiMilla, Paul A},
journal={AIChE J},
volume={44},
number={2},
pages={465--473},
year={1998},
publisher={Wiley Online Library}
}

@article{aben1997photoelastic,
title={Photoelastic tomography for three-dimensional flow birefringence studies},
author={Aben, Hillar and Puro, Alfred},
journal={Inv Probl},
volume={13},
number={2},
pages={215},
year={1997},
publisher={IOP Publishing}
}



\end{document}